\documentclass[aps,prd,twocolumn,superscriptaddress,nofootinbib]{revtex4-1}

\usepackage{aas_macros}
\usepackage[utf8]{inputenc}
\usepackage[T1]{fontenc}
\usepackage[english,french]{babel}
\usepackage{mathrsfs}
\usepackage{amsmath}
\usepackage{amssymb}
\usepackage{tipa}
\usepackage{ntheorem}
\usepackage{braket}
\usepackage{graphicx}
\usepackage{tabularx}
\usepackage{bm}
\usepackage{wasysym}
\usepackage{enumitem}
\usepackage{stmaryrd}
\usepackage[breaklinks=true,colorlinks,citecolor=blue,linkcolor=blue,urlcolor=blue]{hyperref}
\usepackage{tikz}
\usepackage{footmisc}
\usetikzlibrary[patterns]
\usetikzlibrary{shapes}
\usepackage[normalem]{ulem}
\usepackage{multirow}
\usepackage{textcomp}
\usepackage{color}
\usepackage{subcaption}

\theoremseparator{.}

\makeatletter
\newskip\@bigflushglue \@bigflushglue = -100pt plus 1fil

\def\bigcentering{\let\\\@centercr\rightskip\@bigflushglue
\leftskip\@bigflushglue
\parindent\z@\parfillskip\z@skip}

\makeatother

\allowdisplaybreaks[1]

\newcommand{\dd}{\mathrm{d}}

\begin{document}

\title{Detection of magnetic galactic binaries in quasi-circular orbit with LISA}

\author{E. Savalle}
\email{etienne.savalle@cea.fr}
\affiliation{Département de Physique des Particules, CEA/IRFU, CNRS/INSU, Université Paris-Saclay, Université de Paris, 91191 Gif-sur-Yvette, France}

\author{A. Bourgoin}
\email{adrien.bourgoin@obspm.fr}
\affiliation{SYRTE, Observatoire de Paris, Université PSL, CNRS, LNE, Sorbonne Université, 61 avenue de l’Observatoire , 75014 Paris, France}
\affiliation{Université Paris-Saclay, Université Paris Cité, CEA, CNRS, AIM, 91191, Gif-sur-Yvette, France}

\author{C.~Le~Poncin-Lafitte}
\affiliation{SYRTE, Observatoire de Paris, Université PSL, CNRS, LNE, Sorbonne Université, 61 avenue de l’Observatoire , 75014 Paris, France}

\author{S. Mathis}
\affiliation{Université Paris-Saclay, Université Paris Cité, CEA, CNRS, AIM, 91191, Gif-sur-Yvette, France}

\author{M.-C. Angonin}
\affiliation{SYRTE, Observatoire de Paris, Université PSL, CNRS, LNE, Sorbonne Université, 61 avenue de l’Observatoire , 75014 Paris, France}

\author{C. Aykroyd}
\affiliation{SYRTE, Observatoire de Paris, Université PSL, CNRS, LNE, Sorbonne Université, 61 avenue de l’Observatoire , 75014 Paris, France}


\begin{abstract}
  Laser Interferometer Space Antenna (LISA) will observe gravitational waves from galactic binaries (GBs) of white dwarfs or neutron stars. Some of these objects are among the most magnetic astrophysical objects in the Universe. Magnetism, by secularly disrupting the orbit, can eventually affect the gravitational waves emission and could then be potentially detected and characterized after several years of observations by LISA. Currently, the data processing pipeline of the LISA Data Challenge (LDC) for GBs does not consider either magnetism or eccentricity. Recently, it was shown [Bourgoin \emph{et al.} PRD 105, 124042 (2022)] that magnetism induces a shift on the gravitational wave frequencies. Additionally, it was argued that, if the binary's orbit is eccentric, the presence of magnetism could be detected by LISA. In this work, we explore the consequences of a future data analysis conducted on quasi-circular and magnetic GB systems using the current LDC tools. We first show that a single eccentric GB can be interpreted as several GBs and this can eventually bias population studies deduced from LISA's future catalog. Then, we confirm that for quasi-circular orbits, the secular magnetic energy of the system can be inferred if the signal-to-noise ratio of the second harmonic is high enough to be detected by traditional quasi-monochromatic source searching algorithms. LISA observations could therefore bring new insights on the nature and origin of magnetic fields in white dwarfs or neutron stars.
\end{abstract}

\maketitle

\section{Introduction}
\label{sec:int}

Laser Interferometer Space Antenna (LISA) is an ESA (European Space Agency) and NASA (National Aeronautics and Space Administration) joint mission targeted for launch around mid 2030 \cite{LISAcoll,2017arXiv170200786A}. It will consist of three spacecrafts in heliocentric orbits forming a giant space-based triangular interferometer dedicated to the observation of gravitational waves (GWs). With its 2.5 million kilometers arm-length it will watch the low frequency band of the GWs spectrum (i.e., from $0.1\ \mathrm{mHz}$ to $0.1\ \mathrm{Hz}$) hoping to detect the signal from sources which are not yet resolved by the current ground-based detectors such as LIGO \cite{LIGOcoll}, Virgo \cite{Virgocoll}, KAGRA \cite{KRAGAcoll}, GEO600 \cite{GEO600coll}, or the future Einstein Telescope \cite{ETcoll}.

Among the sources that LISA will observe, the galactic binaries (GBs) are composed of white dwarfs (WDs), neutron stars (NSs), and stellar mass black holes in various combinations. It is expected that LISA will be able to resolve more than ten thousand individual GBs over a 5-years LISA mission \cite{PhysRevD.73.122001,2017A&A...602A..16T}. Because of a too low Signal-to-Noise Ratio (SNR) and a large number of individual systems, the rest of the GBs will form a foreground confusion noise in the frequency domain $1\ \mathrm{mHz}$ to $3\ \mathrm{mHz}$ that will not be resolvable by LISA \cite{2010ApJ...717.1006R}. Some of the resolvable GBs are already identified as certified sources of GWs for LISA \cite{Kupfer_2023}. They are called the \emph{verification binaries} and will serve for validating the first run of observations. As a matter of fact, they will constrain the calibration of the detector \cite{Savalle2022} which will impact the bias on other sources.

Currently, the GWs from GBs are modeled assuming a quasi-monochromatic signal \cite{2002ApJ...575.1030T}, namely a signal emitted by a slowly inspiraling binary on circular orbit \cite{LDCGroup}. However, the GWs signal could be more complex than the monochromatic picture, for instance when the orbital motion is not purely circular \cite{2020MNRAS.491.3000M,2021PhRvD.104j4023T} or when it is perturbed by an external process such as a third body perturbation \cite{2007MNRAS.382.1768M,2010MNRAS.407.1048M,2021PhRvD.103f3003W}, or even when the signatures of its own internal interactions such as tidal \cite{2011MNRAS.412.1331F,2012MNRAS.421..426F,2013MNRAS.430..274F,2014MNRAS.444.3488F,2020MNRAS.491.3000M} or magnetic interactions \cite{1990MNRAS.244..731K,PhysRevD.105.124042,2021arXiv210910722M,Bromley_2022} become visible. It is thus important to study the signatures of these perturbations in the GW's signal.

In fact, WDs and NSs are among the most magnetic astrophysical objects of the universe with magnetic fields that can reach up to $10^9\ \mathrm{G}$ for WDs and up to $10^{15}\ \mathrm{G}$ for NSs (i.e., the magnetars) \cite{2020AdSpR..66.1025F}. These fields are so intense that they could significantly change the orbital motion of a GB and then generate detectable signatures in the GWs signal \cite{PhysRevD.105.124042,2022GReGr..54..146L}. Hence, LISA could measure the magnetism within thousands of binary systems and could thus be the opportunity to learn more on the origin and the nature of magnetism in degenerate stars by complementing observations from large spectroscopic surveys \cite{2013MNRAS.429.2934K,2015MNRAS.450..681H} and from polarimetric surveys \cite{2012A&A...545A..30L,2018A&A...618A.113B,2022ApJ...935L..12B}. 

The observed diversity in the characteristics of magnetic fields in degenerate stars (and in main sequence stars too) has led astrophysicists to come up with different scenarios of formation for magnetic WDs and NSs. Schematically, three different channels have been proposed, namely the ``merging scenario'', the ``dynamo hypothesis'', and the ``fossil field'' hypothesis.

\begin{enumerate}
  \item The merging scenario states that magnetic fields are amplified by a dynamo during accretion of rocky debris \cite{2011MNRAS.413.2559F,2018MNRAS.478..899B} or during the merger of a binary pair \cite{2005MNRAS.356..615F,2008MNRAS.387..897T}. Originally, it was motivated by the dearth of magnetic WDs in detached binary systems \cite{2005AJ....129.2376L,Olausen_2014}. However, as pointed out by \citet{2016Natur.537..374M} and \citet{2020A&A...634L..10L}, strongly magnetic WDs in binary systems actually do exist. Furthermore, \citet{2021MNRAS.507.5902B} concluded after analyzing spectropolarimetric observations of more than 150 WDs within $20\ \mathrm{pc}$ from the Sun, that there is no strong indication that the frequency of isolated single magnetic WDs is different than in binary systems.
  \item The dynamo hypothesis encompasses scenarios where the magnetic field is generated after the collapse of WD/NS's progenitors. The NS's version of this scenraio involves a dynamo right after the mass ejection of the progenitor where a vigorous convective episode together with a fast rotation could give rise to a strong magnetic field on a short period of time (few seconds) \cite{1992ApJ...392L...9D,2020SciA....6.2732R,2022A&A...667A..94R}. Even if this scenario seems to recover some properties of millisecond pulsars it still faces the fact that the current population of magnetars seems to favor slow rotators \cite{Olausen_2014}. The WD's version of the dynamo hypothesis suggests that during the cooling, the strong convection induced by the crystallisation of WDs' C-O core together with a rapid rotation-driven dynamo could generate a strong magnetic field that would then diffuse from the inner region after billion years \cite{2017ApJ...836L..28I,2021NatAs...5..648S}. This scenario, by favoring old WDs as natural hosts for magnetic fields, is supported by the dearth of young magnetic WDs (with ages below $1\ \mathrm{Gyr}$) observed by \citet{2021MNRAS.507.5902B}.
  \item The fossil-fields hypothesis suggests that magnetic fields in compact stars are inherited from pre-WD/NS's evolutions. In a first version, the fields are induced by flux conservation during evolution from the main sequence to the final compact state \cite{1964ApJ...140.1309W,1967PhRv..153.1372L,1981ApJS...45..457A,2005MNRAS.356..615F}. Accordingly, the progenitors of strongly magnetic WDs would be the Ap and Bp stars while those of NSs would be the stars of spectral type O with strong effective dipolar fields \cite{2005MNRAS.356..615F,2009MNRAS.396..878H}. However, as initially pointed out by \citet{2007ApJ...654..499K} and confirmed later on by \citet{2021MNRAS.507.5902B}, this channel is probably not the main one to produce the 20\% of magnetic WDs from the 8\% of A and B main sequence stars \cite{2008CoSka..38..443P}. A second version of the fossil-fields hypothesis states that a dynamo-driven magnetic field would be generated in the convective cores of stars either in the main sequence or in the Asymptotic Giant Branch \cite{2005ApJ...629..461B,2006MNRAS.370..629D,2010ApJ...724L..34D,2010A&A...517A..58D}. It would then be compressed and amplified during the latter evolution, and then revealed after the ejection of the envelope. Finally, the diffusion of the field from the inner compact star to its surface usually unveils a stable dipolar magnetic field \cite{2004Natur.431..819B,2008MNRAS.386.1947B}. For slowly diffusing fields, this mechanism is compatible with the dearth of young magnetic WDs observed by \citet{2021MNRAS.507.5902B}.
\end{enumerate}

This non-exhaustive list of possible channels of formation of magnetic fields illustrates the fact that additional data are needed in order to refine the scenario of formation. By observing tens of thousands of GBs, LISA represents the perfect opportunity to further constrain the nature of magnetism within WDs and NSs. In this attempt, the quasi-monochromatic picture is likely not accurate enough over a 5-years LISA mission, and it can even bias the calibration of the detector or the determination of the most sensitive physical parameters (e.g., the chirp mass, etc.). Therefore, the data processing pipelines have to be refined in order to fully extract the physical information from the data. As a matter of fact, it was shown recently \cite{PhysRevD.105.124042} that the possibility of detecting magnetism from the GWs signal alone is intrinsically linked to the presence of eccentricity in the binary system. However, this cannot be tested within the monochromatic picture of the current data processing pipeline of the LISA Data Challenge (LDC)  algorithms. The magnetic dipole-dipole interaction generates a secular rate of change in the mean longitude and in the longitude of the pericenter. These secular drifts are proportional to the secular magnetic energy of the binary system. For a binary in circular orbit, the frequency of the GWs signal is proportional to the mean longitude, thus magnetism shifts the frequency of the GWs signal with respect to the non-magnetic case. This shift is fully degenerated with the frequency and hence it should not be detectable from GWs observations alone. However, the situation slightly changes if the source of the signal is in quasi-circular orbit. Indeed, in this configuration, the GWs signal is more complex than the monochromatic picture since it contains additional frequencies besides the main frequency of the circular orbit. Magnetism will then shift all the frequencies, so that the magnetic information can actually be deduced from the GWs observations by measuring and combining the harmonic frequencies. Therefore, the possibility of observing magnetism from GWs is fully related to the possibility of detecting magnetic binary systems in quasi-circular orbit. If this is challenging according to the merging scenario it is perfectly feasible within the fossil-fields hypothesis or the dynamo scenario.

In this work, we address the problem of determining whether or not current LDC monochromatic algorithms are sufficiently robust to allow for the detection of all GBs especially those which are in quasi-circular orbit and which host magnetic interactions. Then, we derive a simplified procedure for determining the magnetic information of the binary system from the determination of the main frequency and second harmonic of the GW signal. The paper is organized as follows. In Sect. \ref{sec:modepol}, we recall the necessary results of \citet{PhysRevD.105.124042} concerning the secular orbital dynamics of a binary system in quasi-circular orbit while considering the effects of general relativity (GR) and of the dipole-dipole magnetic interaction. In Sect. \ref{sec:waveform}, we derive the corresponding GWs waveform up to the second-order in eccentricity and then we relate the measurable quantities (e.g., the amplitudes or the frequencies of the signal) to the physical parameters of the binary system (e.g., the chirp mass or the product of the magnetic moments). In Sect. \ref{sec:numerical_setup}, we use the computed waveform to simulate the LISA data and then we analyse it assuming the quasi-monochromatic picture from the LDC. The GW signal is simulated using the parameters from an identified verification binary, namely HM Cancri - RX J0806.3+1527, whose eccentricity could be of the order of 0.1 \cite{2020MNRAS.491.3000M}. In Sect. \ref{sec:result}, we discuss the outcome (i.e., the inference of the GWs signal's parameters) of the data analysis considering a 1, 4, and 8-years LISA mission. We then determine the estimates of the physical parameters thanks to the simplified expressions determined previously in Sect. \ref{sec:waveform}. Finally, we give our conclusions in Sect. \ref{sec:ccl}. 

\section{Orbital dynamics of magnetic Galactic Binaries}
\label{sec:modepol}

In order to analyze the GWs signal from a particular source, we first need to describe its orbital motion. We assume a binary system in quasi-circular orbit (eccentricity $\ll 1$) and consider the effects of (i) GR up to the 2.5 Post-Newtonian (PN) approximation and (ii) the magnetic dipole-dipole interaction. The description of the secular motion has been derived in \citet{PhysRevD.105.124042}. In this section, we recall the main results that are necessary for our purpose.

\subsection{Reference frames}

In this work, we suppose that the observer (i.e., the LISA constellation) is in the far-away wave-zone \cite{2014grav.book.....P} so that we call it the \emph{far-away observer}. Hence, the field point (i.e., the LISA constellation) is considered far from the source point (i.e., the center-of-mass of the binary system) in the sense that the separation between the two is much larger than the characteristic wavelength of the GWs emitted by the source (i.e., the magnetic GB). Accordingly, the relative motion of the observer can be neglected and $\hat{\mathbf{N}}$, the unit-direction of the observer relative to the source point, can be considered constant.

Following notations in \cite{PhysRevD.105.124042}, let $(\hat{\mathbf{e}}_X,\hat{\mathbf{e}}_Y,\hat{\mathbf{e}}_Z)$ be the orthonormal basis attached to the center-of-mass of the binary system; we shall call this frame the \emph{source frame}. By definition, the source frame is an inertial frame and is thus well-suited for the description of the orbital motion of the binary system (see Sect.~\ref{sec:orbmot}). The choice of the orientation of the source frame will be discussed in more depth in Sect. \ref{sec:waveform}.

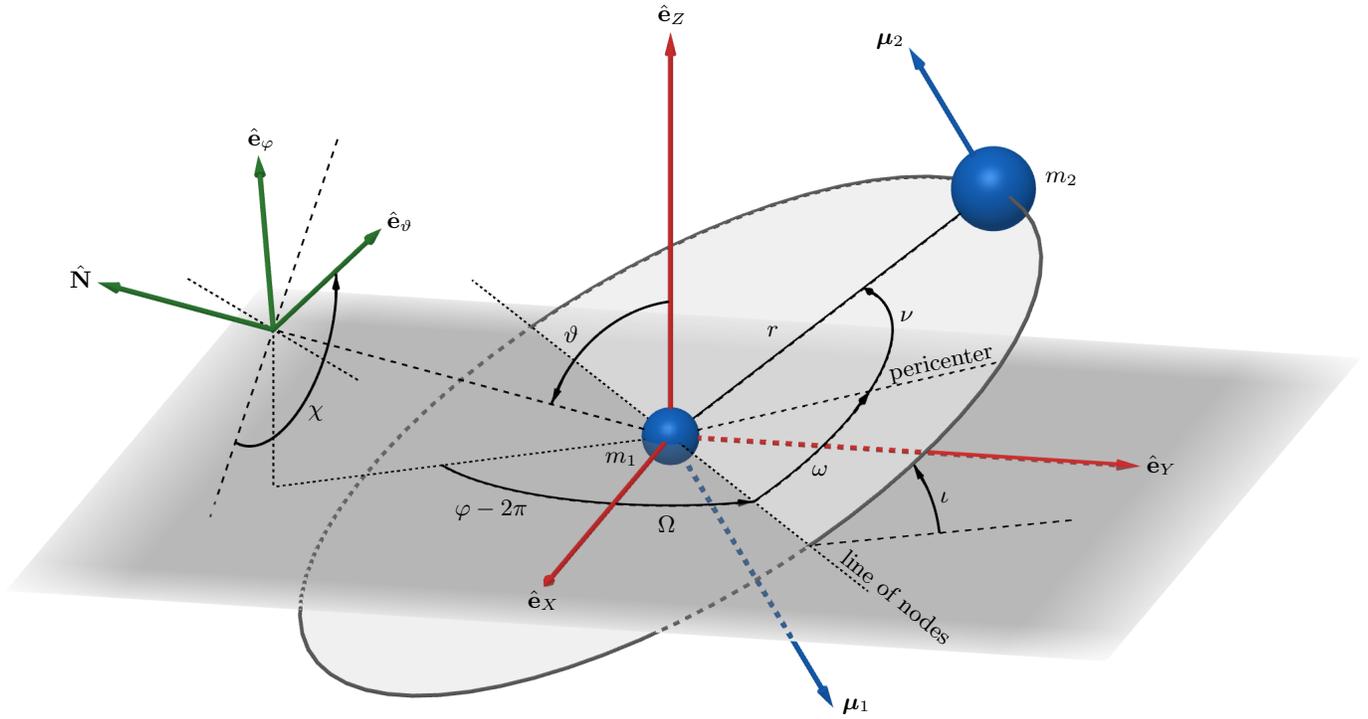
\begin{figure*}
  \begin{center}
    \includegraphics[scale=0.25]{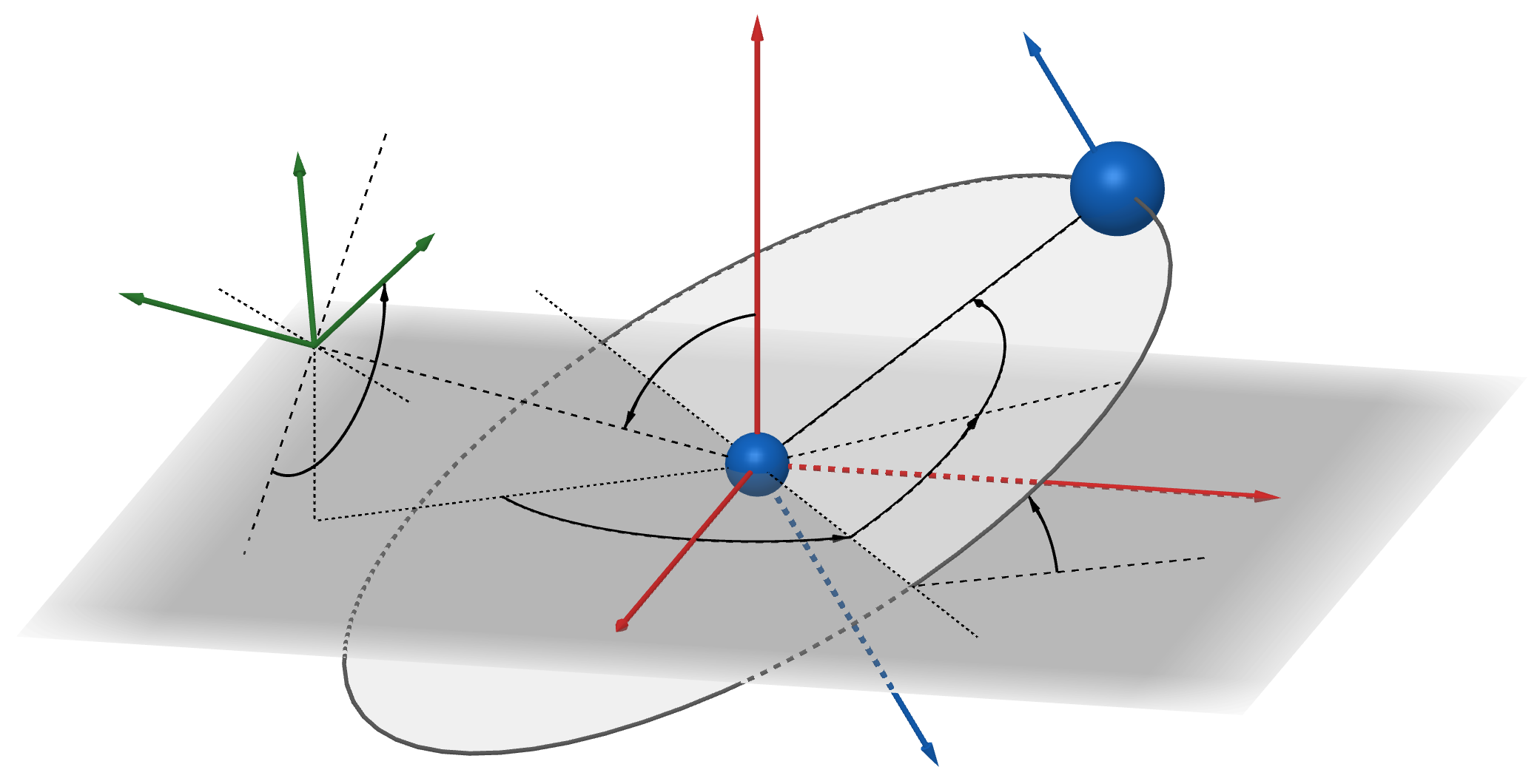}
  \end{center}
  \setlength{\unitlength}{1.0cm}
  \begin{picture}(0,0)
    \put(-1.83,2.25){\rotatebox{0}{$\hat{\mathbf e}_X$}}
    \put(6.4,4.05){\rotatebox{0}{$\hat{\mathbf e}_Y$}}
    \put(-0.1,10.05){\rotatebox{0}{$\hat{\mathbf e}_Z$}}
    \put(-3.7,7.3){\rotatebox{0}{$\hat{\mathbf e}_\vartheta$}}
    \put(-5.55,8.4){\rotatebox{0}{$\hat{\mathbf e}_\varphi$}}
    \put(-7.92,6.5){\rotatebox{0}{$\hat{\mathbf N}$}}
    \put(-0.8,4.17){\rotatebox{0}{$m_1$}}
    \put(5.05,7.9){\rotatebox{0}{$m_2$}}
    \put(1.35,5.85){\rotatebox{0}{$r$}}
    \put(-1.35,5.78){\rotatebox{0}{$\vartheta$}}
    \put(-2.8,3.47){\rotatebox{0}{$\varphi-2\pi$}}
    \put(-4.75,4.8){\rotatebox{0}{$\chi$}}    
    \put(-0.1,3.25){\rotatebox{0}{$\Omega$}}
    \put(1.95,4.0){\rotatebox{0}{$\omega$}}
    \put(3.65,3.65){\rotatebox{0}{$\iota$}}
    \put(3.12,6.05){\rotatebox{0}{$\nu$}}
    \put(2.31,2.87){\rotatebox{-39.}{$\text{line of nodes}$}}
    \put(2.95,5.28){\rotatebox{12}{$\text{pericenter}$}}
    \put(2.8,9.75){\rotatebox{0}{$\bm{\mu}_2$}}
    \put(2.35,0.9){\rotatebox{0}{$\bm{\mu}_1$}}
  \end{picture}
  \caption{Orientation of $(\hat{\mathbf e}_\vartheta,\hat{\mathbf e}_\varphi,\hat{\mathbf N})$, the transverse frame (\emph{in green}), in the source frame (\emph{in red}), namely $(\hat{\mathbf e}_X,\hat{\mathbf e}_Y,\hat{\mathbf e}_Z)$. The primary is shown at the center-of-mass of the binary system in order to simplify the drawing -- this corresponds to the case where the mass of the secondary is negligible with respect to primary's. The orbital motion of the binary system is described at any time $t$ by making use of the radial separation $(r)$, the inclination $(\iota)$, the longitude of the node $(\Omega)$, the argument of the pericenter $(\omega)$, and the true anomaly $(\nu)$. The orientation of the transverse frame with respect to the source frame is parameterized by the following triple of angles: $(\vartheta,\varphi,\chi)$. The magnetic moments of the primary and secondary (\emph{in blue}) are denoted by $\bm\mu_1$ and $\bm\mu_2$, respectively.}
  \label{fig:orbit}
\end{figure*}

Finally, let $(\hat{\mathbf{e}}_\vartheta,\hat{\mathbf{e}}_\varphi,\hat{\mathbf N})$ be the orthonormal basis associated to the far-away observer; we shall call this frame the \emph{transverse frame}. The denomination is motivated by the fact that in the far-away wave-zone, the harmonic gauge can be further specialized by using the transverse-tracefree gauge \cite{2014grav.book.....P}. The orientation of the transverse frame in the source frame is represented in figure \ref{fig:orbit}.

\subsection{Secular orbital dynamics}
\label{sec:orbmot}

The emission of GWs is known to induce a back reaction on the orbit making it more and more circular. Therefore, it is expected that, after a certain period of time $t_{\mathrm{GW}}$, an eccentric binary system will reach a quasi-circular orbit: that is to say $e\ll 1$ with $e$ being the eccentricity. The characteristic time for circularizing the orbit is typically on the order of $t_{\mathrm{GW}}\sim c^5(Gm)^{-5/3}f^{-8/3}$ with $c$ the speed of light in a vacuum, $G$ the gravitational constant, $f$ the main frequency of the gravitational radiation, and $m$ the total mass of the binary system namely $m=m_1+m_2$ with $m_1$ and $m_2$ the masses of primary and secondary, respectively. For typical GBs in the high frequency band of LISA (i.e., for $f=0.1\ \mathrm{Hz}$), the characteristic time for the GWs radiation is about $t_{\mathrm{GW}}\sim 10^4\ \mathrm{yr}$. The loss of eccentricity can thus occur on relatively short time scales with respect to the age of certain binary systems \cite{2021MNRAS.507.5902B}.

For small values of the eccentricity, the orbital motion is conveniently described by making use of the regular orbital elements, namely $a$, $z$, $\zeta$ with $a$ the semi-major axis, $z$ the imaginary eccentricity vector, $\zeta$ the imaginary inclination vector, and $L$ the mean longitude of the orbit. The expressions of the regular elements are as follows:
\begin{subequations}\label{eq:defregparam}
  \begin{align}
    z&=e\,\mathrm{exp}\left(\mathrm i\varpi\right)\text{,}\\
    \zeta&=\sin\left(\frac{\iota}{2}\right)\mathrm{exp}\left(\mathrm i\Omega\right)\text{,}\\
    L&=\varpi+M\text{.}
  \end{align}
\end{subequations}
The imaginary number is $\mathrm{i}\equiv\sqrt{-1}$ and $\iota$ is the inclination of the orbital plane on the inertial equator of the source frame $(\hat{\mathbf{e}}_X,\hat{\mathbf{e}}_Y)$. $\Omega$ is the longitude of the ascending node and $\varpi=\Omega+\omega$ is the longitude of the pericenter with $\omega$ the argument of the pericenter. $M$ is the mean anomaly which is expressed as $M=n(t-\tau)$, where $\tau$ is the time of pericenter passage with $n$ the mean motion being given by the Kepler's third law: 
\begin{equation}
  n=\sqrt{\frac{Gm}{a^3}}\text{.}
  \label{eq:kep3}
\end{equation}
The different angular elements are represented in Fig. \ref{fig:orbit}.

When the binary system dynamics is only driven by the Newton two-body equations of motion, the regular elements are constants, apart from the mean longitude which evolves linearly in time, namely $L(t)\propto nt$ with $n$ being a constant. However, when considering perturbations to the Newton two-body equations of motion, we expect the regular elements to evolve in time. If the magnitude of the perturbation is small with respect to the two-body zeroth-order term, the characteristic time for the change in the regular elements is expected to be much larger than one orbital period $P=2\pi/n$. In this case, it is convenient to focus on the long-term variations of the regular elements, namely their secular variations.

Let us now report the secular dynamics of a magnetic GB within the LISA frequency band. As mentioned earlier, the system is thus expected to be in the inspiral phase and in quasi-circular orbit due to the gravitational radiation. By considering the perturbations from GR (up to the 2.5PN approximation) and the magnetic dipole-dipole interaction (on both the orbital and rotational motions), we can derive the approximate secular evolution of the regular elements. We thus keep up to the quadratic terms in time and systematically neglect the periodic ones which all have small amplitudes in the present context of an inspiral magnetic binary system as discussed in \citet{PhysRevD.105.124042}. Hence, the secular evolution of the longitude of the pericenter and the mean longitude are approximated by
\begin{subequations}\label{eq:solpiL}
\begin{align}
  \varpi(t)&=\varpi_0+(\dot\varpi_{1\mathrm{PN}}+\dot\varpi_{\mathrm{M}})t\text{,}\\
  L(t)&=L_0+(n_0+\dot L_{1\mathrm{PN}}+\dot L_{\mathrm{M}})t+\frac{3n_0}{4}\frac{|\dot a_{2.5\mathrm{PN}}|}{a_0}t^2\text{,}\label{eq:dotLGR}
\end{align}
\end{subequations}
where an ``overdot'' denotes a rate of change. The secular contribution from GR is denoted by a subscript ``$x$PN'' where ``$x$'' refers to the degree of the post-Newtonian expansion. The secular contribution from magnetism is denoted by a subscript ``M''. In the previous expressions and hereafter, we systematically neglect the 2PN terms before the 1PN contribution. The expressions of the 1PN terms are given by
\begin{subequations}\label{eq:solpiLGR}
\begin{align}
  \dot\varpi_{1\mathrm{PN}}&=3n_0{}^{5/3}\left(\frac{G\mathcal M}{c^3}\right)^{2/3}\frac{\eta^{-2/5}}{(1-e_0{}^2)}\text{,}\\
  \dot L_{1\mathrm{PN}}&=\dot\varpi_{1\mathrm{PN}}\Bigg\{\frac{10}{3}\left(1-\frac{\eta}{2}\right)\frac{1-\sqrt{1-{e_0}^2}}{{e_0}^2}\nonumber\\
  &-\frac{5}{3}\sqrt{1-{e_0}^2}-\frac{\eta}{6}\left(1-8\sqrt{1-{e_0}^2}\right)\nonumber\\
  &-\frac{7}{3}{e_0}^2\left(1-\frac{11\eta}{14}\right)\Bigg\}\text{,}
\end{align}
\end{subequations}
where $n_0=n(t=0)$, $a_0=a(t=0)$, and $e_0=e(t=0)$ are the initial mean motion, the initial semi-major axis, and the initial eccentricity, respectively. The symmetric mass ratio and the chirp mass, namely $\eta$ and $\mathcal{M}$, are given respectively by
\begin{equation}
  \eta=\frac{m_1m_2}{m^2}\text{,}\qquad \mathcal{M}=\eta^{3/5}m\text{.}
  \label{eq:chirp}
\end{equation}

The secular magnetic terms read as follows:
\begin{subequations}
\begin{align}
  \dot\varpi_{\mathrm{M}}&=\frac{3\bar U_{\mathrm{M}}}{\sqrt{1-{e_0}^2}}\left(\frac{n_0}{G^2\mathcal{M}^5}\right)^{1/3}\text{,}\label{eq:omegamag}\\
  \dot L_{\mathrm{M}}&=\dot\varpi_{\mathrm{M}}\left(1+\sqrt{1-{e_0}^2}\right)\text{,}\label{eq:Lmag}
\end{align}
\end{subequations}
where $\bar U_{\mathrm{M}}$ is a constant quantity representing the secular magnetic energy of the binary system, namely
\begin{equation}
  \bar{U}_{\mathrm{M}}=\frac{\mu_0}{4\pi a_0{}^3(1-e_0{}^2)^{3/2}}\big(3\mu_1^z\mu_2^z-\bm{\mu}_1\cdot\bm{\mu}_2\big).
  \label{eq:magE}
\end{equation}
In this last expression, $\mu_0$ is the vacuum magnetic permeability, $\bm{\mu}_1$ and $\bm{\mu}_2$ are the dipole magnetic moments of the primary and secondary, respectively, and $\mu_1^z$ and $\mu_2^z$ are the projection of $\bm{\mu}_1$ and $\bm{\mu}_2$ on the direction of the orbital angular momentum $\hat{\mathbf{e}}_z$, respectively.

The solutions for the secular evolution of the semi-major axis and eccentricity vector are given by
\begin{subequations}\label{eq:solaz}
\begin{align}
  a(t)&=a_0-|\dot a_{2.5\mathrm{PN}}|t\text{,}\\
  z(t)&=e_0\,\mathrm{exp}\left(-\frac{|\dot e_{2.5\mathrm{PN}}|}{e_0}t\right)\mathrm{exp}\big(\mathrm{i}\varpi(t)\big)\text{,}
\end{align}
\end{subequations}
where the GR contributions, at the 2.5PN approximation, are given by
\begin{subequations}
\begin{align}
  \frac{\dot a_{2.5\mathrm{PN}}}{a_0}&=-\frac{64}{5}n_0{}^{8/3}\bigg(\frac{G \mathcal{M}}{c^3}\bigg)^{5/3}\frac{\mathcal{G}(e_0)}{\left(1-e_0{}^2\right)^{7/2}}\text{,}\\
  \frac{\dot e_{2.5\mathrm{PN}}}{e_0}&=-\frac{304}{15}n_0{}^{8/3}\bigg(\frac{G\mathcal{M}}{c^3}\bigg)^{5/3}\frac{\mathcal{H}(e_0)}{\left(1-e_0{}^2\right)^{5/2}}
\end{align}
\end{subequations}
with $\mathcal{G}(e_0)$ and $\mathcal{H}(e_0)$ being given by
\begin{subequations}
\begin{align}
  \mathcal{G}(e_0)&=1+\frac{73}{24}e_0{}^2+\frac{37}{94}e_0{}^4\text{,}\\
  \mathcal{H}(e_0)&=1+\frac{121}{304}{e_0}^2\text{.}
\end{align}
\end{subequations}

Neither GR nor the magnetic dipole-dipole interaction can secularly change the inclination vector, hence
\begin{equation}
  \zeta(t)=\zeta_0\text{.}
  \label{eq:zeta}
\end{equation}
Furthermore, one might see that the magnetic interaction, in the magnetostatic approximation, does not secularly affect the shape of the orbit, namely $a$ and $e$. The secular variations of the semi-major axis and eccentricity are caused by the loss of orbital energy and angular momentum that are carried away from the binary system by the gravitational radiations (i.e., the 2.5PN terms). This yields a decrease in semi-major axis and to an exponential decay in eccentricity, namely an efficient circularization of the orbit. The linear drift of the semi-major axis causes an increase of the orbital frequency (i.e., the mean motion) and this makes the mean longitude vary quadratically in time as shown in Eq.~\eqref{eq:dotLGR}.

From the orbital motion, we can now determine the mode polarizations of the GWs. This is the subject of the next section.

\section{Gravitational waveform}
\label{sec:waveform}

For an observer in the far-away wave-zone, it is well-known \cite{1963PhRv..131..435P,2014grav.book.....P} that the GW mode polarizations $h_+$ and $h_\times$ are the transverse-tracefree part of the potentials $h^{jk}$ (which is expressed in the source frame), that is to say
\begin{subequations}\label{eq:GWmodes}
\begin{align}
  h_{+}&=\tfrac{1}{2}\big[(\hat{\mathbf e}_\vartheta)_j(\hat{\mathbf e}_\vartheta)_k-(\hat{\mathbf e}_\varphi)_j(\hat{\mathbf e}_\varphi)_k\big]h^{jk}\text{,}\\
  h_{\times}&=\tfrac{1}{2}\big[(\hat{\mathbf e}_\vartheta)_j(\hat{\mathbf e}_\varphi)_k+(\hat{\mathbf e}_\varphi)_j(\hat{\mathbf e}_\vartheta)_k\big]h^{jk}\text{.}  
\end{align}
\end{subequations}
The components $h^{jk}$ are given by the quadrupole formula
\begin{equation}
  h^{jk}(t_{\mathrm{obs}},\mathbf x)=\frac{2G}{c^4D}\,\ddot{I}^{jk}(t)\text{.}
  \label{eq:potquad}
\end{equation}
In this expression, $D$ is the separation between the source and the field point, namely $D=\Vert\mathbf x\Vert$, and $t$ is the retarded time, namely $t=t_{\mathrm{obs}}-D/c$. Finally, $I^{jk}$'s are the components of the quadrupole moment of inertia expressed in the source frame, namely
\begin{equation}
  I^{jk}(t)=\int\rho(t,\mathbf x)\,{x}^j{x}^k\,\dd^3 x\text{,}
  \label{eq:quadInert}
\end{equation}
with $\rho$ the density matter distribution.

For a binary system composed of two punctual masses, the expression of the second time derivative of \eqref{eq:quadInert} can be straightforwardly derived from the tensor virial theorem. It involves the well-known solution of the Kepler motion. Therefore, the mode polarizations \eqref{eq:GWmodes} are conveniently expressed in term of the orbital elements $(a,z,\zeta,L)$, where the mean longitude is the only angle varying on short timescale, namely one orbital period \cite{1963PhRv..131..435P,2012ApJ...752...67K}. In this picture, the final relationships of $h_+$ and $h_\times$ are valid for circular and eccentric orbits.

When keeping the transverse-tracefree part of $h^{jk}$, the mode polarizations depend on the orientation of the transverse frame in the source frame through the angles $\vartheta$, $\varphi$, and $\chi$ which are constants for an observer in the far-away wave zone. Therefore, specifying the orientation of the source frame allows one to remove unnecessary constants. From now, we assume that the source frame coincides with the transverse frame such that $(\hat{\mathbf{e}}_X,\hat{\mathbf{e}}_Y,\hat{\mathbf{e}}_Z)$ $\rightarrow$ $(\hat{\mathbf{e}}_{\vartheta},\hat{\mathbf{e}}_{\varphi},\hat{\mathbf{N}})$, and this fixes all the degrees of freedom in the orientation of the source frame besides a remaining arbitrary rotation around $\hat{\mathbf{e}}_Z$ (i.e., $\hat{\mathbf{N}}$). As it might be seen from figure \ref{fig:orbit}, this transformation can be achieved by substituting $\vartheta=\varphi=0$ and $\chi=0$ into the expressions for the mode polarizations.

\subsection{Fourier decomposition of the waveform}

Having chosen the orientation of the source frame, the mode polarizations $h_+$ and $h_\times$ are now given at the $\ell$th-order in $|z|$ (i.e., up to the $\ell$th-order in eccentricity) by the following Fourier series:
\begin{equation}
  h_+-\mathrm{i}h_\times=h(a)\sum_{k=-(\ell+2)}^{\ell+2}c_k(z,\zeta)\,\mathrm{e}^{\mathrm i kL}\text{,}
  \label{eq:h+hx}
\end{equation}
where the Fourier's coefficients $c_k$ are given hereafter up to second-order in eccentricity (i.e., $\ell=2$):
\begin{subequations}\label{eq:FourierCoef}
\begin{align}
  c_{+4}&=-8\bar z^2\bar\zeta^4\text{,}\\
  c_{+3}&=-\tfrac{9}{2}\bar z\bar\zeta^4\text{,}\\
  c_{+2}&=-2\left(1-\tfrac{5}{2}z\bar z\right)\bar\zeta^4+\bar z^2\bar\zeta^2\left(1-\zeta\bar\zeta\right)\text{,}\\
  c_{+1}&=\tfrac{3}{2}z\bar\zeta^4+\bar z\bar\zeta^2\left(1-\zeta\bar\zeta\right)\text{,}\\
  c_{-1}&=\tfrac{3}{2}\bar z\left(1-\zeta\bar\zeta\right){}^2+z\bar\zeta^2\left(1-\zeta\bar\zeta\right)\text{,}\\
  c_{-2}&=-2\left(1-\tfrac{5}{2}z\bar z\right)\left(1-\zeta\bar\zeta\right){}^2+z^2\bar\zeta^2\left(1-\zeta\bar\zeta\right)\text{,}\\
  c_{-3}&=-\tfrac{9}{2}z\left(1-\zeta\bar\zeta\right){}^2\text{,}\\
  c_{-4}&=-8z^2(1-\zeta\bar\zeta)^2\text{,}
\end{align}
\end{subequations}
with $c_0=0$. Let us recall that the complex variables $z$ and $\zeta$ are defined in Eqs. \eqref{eq:defregparam}. The GW strain amplitude, $h$, is function of the mean motion [or equivalently the semi-major axis, see Eq. \eqref{eq:kep3}] and is given by
\begin{equation}
  h(a)=\frac{2c\,n^{2/3}}{D}\left(\frac{G\mathcal M}{c^3}\right)^{5/3}\text{.}
  \label{eq:amp}
\end{equation}

Equation \eqref{eq:h+hx} is still valid beyond the Kepler motion. Hence, by substituting the secular solutions \eqref{eq:solaz} and \eqref{eq:solpiL} into the right-hand side of Eq.~\eqref{eq:h+hx}, one can get the combined effects of GR (up to the 2.5PN order) and the magnetic dipole-dipole interaction on the mode polarizations. Then, according to Eq. \eqref{eq:kep3}, we have the relation: $n^{2/3}(t)\propto a^{-1}(t)$, and hence the GW strain amplitude increases as the semi-major axis decreases because of the energy loss due to the gravitational radiation. On the other hand, the magnetic dipole-dipole interaction (and the 1PN terms) secularly changes the mean longitude and the longitude of the pericenter. Because the latter enters into Eq. \eqref{eq:h+hx} at first-order in eccentricity, it can be neglected for quasi-circular orbits and only the secular drift of the mean longitude needs to be considered. Therefore, we anticipate that magnetism will slightly change the frequency of the mode polarizations with respect to the frequency that would be expected for two point-masses orbiting each other on circular orbits.

\subsection{The detector adapted frame}\label{sec:detectorframe}

Following \citet{2014grav.book.....P}, we further specialize the source frame by imposing that $\hat{\mathbf{e}}_X$ (i.e., $\hat{\mathbf{e}}_\vartheta$) is aligned with the direction of the line of nodes. In other words, we now set $\Omega=0$, which according to Eq. \eqref{eq:defregparam}, leads to $\zeta=\sin({\iota/2})$. This is a convenient convention to choose since neither GR nor the magnetic dipole-dipole interaction do change the longitude of the ascending node [see Eq. \eqref{eq:zeta}]. The source frame now coincides exactly with the LDC conventions \cite{LDCGroup} and with the \emph{detector adapted} frame introduced in \citet{2014grav.book.....P}.

The Fourier series \eqref{eq:h+hx} can thus be simplified and expressed as a trigonometric expansion where the ``$+$'' and ``$\times$'' polarizations are now separated:
\begin{widetext}
\begin{subequations}
\begin{align}
  \frac{h_{+}}{h(a)}&=-\left(1-\tfrac{5}{2}e^2\right)(1+\cos^2\iota)\cos 2L-e\Big\{(1+\cos^2\iota)\Big[\tfrac{9}{4}\cos(3L-\varpi)-\tfrac{3}{4}\cos(L+\varpi)\Big]-\tfrac{1}{2}\sin^2\iota\cos(L-\varpi)\Big\}\nonumber\\
  &-e^2\Big\{4(1+\cos^2\iota)\cos(4L-2\varpi)-\tfrac{1}{2}\sin^2\iota\cos(2L-2\varpi)\Big\}\text{,}\\
  \frac{h_{\times}}{h(a)}&=-2\left(1-\tfrac{5}{2}e^2\right)\cos\iota\sin 2L-e\cos\iota\Big[\tfrac{9}{2}\sin(3L-\varpi)-\tfrac{3}{2}\sin(L+\varpi)\Big]-8e^2\cos\iota\sin(4L-2\varpi)\text{.}
\end{align}
\label{eq:hphc}
\end{subequations}

\end{widetext}
In these expressions, we neglected third-order terms in eccentricity. These expressions are used in Sect. \ref{sec:numerical_setup} to simulate the gravitational signal from a verification binary called HM Cancri, but before let us investigate the impact of the orbital dynamics on the GW signal when considering GR and the magnetic dipole-dipole interactions.

\subsection{Frequency spectrum of a magnetic GB in quasi-circular orbit}
\label{sec:phys_inter}

In this section, we focus on the dominant contribution right after the main peak at $2L$ which is the only one remaining for the special case of purely circular orbits. Obviously the same reasoning as the one presented hereafter would also apply to investigate higher order harmonics.

After substituting the secular solutions \eqref{eq:solpiL} and \eqref{eq:solaz} into Eqs. \eqref{eq:hphc}, we obtain the following expressions given here up to the first-order in eccentricity (the number between parenthesis represents the order in eccentricity)
\begin{subequations}
\begin{align}
  h_+(t)&=h_+^{(0)}(t)+h_+^{(1)}(t)\text{,}\\ h_\times(t)&=h_\times^{(0)}(t)+h_\times^{(1)}(t)\text{,}
\end{align}
\end{subequations}
where the zeroth-order terms are
\begin{subequations}\label{eq:hpc1}
  \begin{align}
    h_{+}^{(0)}(t)&=\mathcal{A}_{(0)}(t)(1+\cos^2\iota)\cos\Phi_{(0)}(t)\text{,}\\
    h_{\times}^{(0)}(t)&=2\mathcal{A}_{(0)}(t)\cos\iota\sin\Phi_{(0)}(t)\text{.}
  \end{align}  
\end{subequations}
The amplitude $\mathcal{A}_{(0)}$ and the phase $\Phi_{(0)}(t)$ are respectively given by 
\begin{subequations}
\begin{align}
  \mathcal{A}_{(0)}(t)&=-h_0\left[\frac{1-\frac{5}{2}e_0{}^2\,\mathrm{exp}\left(-2\frac{|\dot{e}_{2.5\mathrm{PN}}|}{e_0}t\right)}{1-\frac{|\dot{a}_{2.5\mathrm{PN}}|}{a_0}t}\right]\text{,}\label{eq:Amp0}\\
  \Phi_{(0)}(t)&=2\pi f_{(0)}t+\pi\dot f_{(0)}t^2-\phi_{(0)}\text{,}   
\end{align}
\end{subequations}
with $\phi_{(0)}=-2L_0$ and $h_0=h(a_0)$ [cf. Eq. \eqref{eq:amp}]. The main frequency and the time frequency shift are respectively defined by
\begin{subequations}\label{eq:fdotf0}
\begin{align}
  f_{(0)}&=\frac{n_0}{\pi}\bigg(1+\frac{\dot L_{1\mathrm{PN}}}{n_0}+\frac{\dot L_{\mathrm{M}}}{n_0}\bigg)\text{,}\label{eq:f0}\\
  \dot f_{(0)}&=\frac{3n_0}{2\pi}\frac{|\dot a_{2.5\mathrm{PN}}|}{a_0}\text{.}
\end{align}
\end{subequations}

The expressions for the the mode polarizations at first-order in eccentricity are
\begin{subequations}\label{eq:h1ecc}
  \begin{align}
    h_{+}^{(1)}(t)&=\mathcal{A}_{(1)}(t)(1+\cos^2\iota)\cos\Phi_{(1)}(t)+\ldots\text{,}\\
    h_{\times}^{(1)}(t)&=2\mathcal{A}_{(1)}(t)\cos\iota\sin\Phi_{(1)}(t)+\ldots\text{,}
  \end{align}  
\end{subequations}
where the amplitude $\mathcal{A}_{(1)}$ and the phase $\Phi_{(1)}(t)$ are respectively given by
\begin{subequations}
  \begin{align}
    \mathcal{A}_{(1)}(t)&=-\frac{9e_0h_0}{4}\left[\frac{\mathrm{exp}\left(-\frac{|\dot{e}_{2.5\mathrm{PN}}|}{e_0}t\right)}{1-\frac{|\dot{a}_{2.5\mathrm{PN}}|}{a_0}t}\right]\text{,}\label{eq:Amp1}\\
    \Phi_{(1)}(t)&=2\pi f_{(1)}t+\pi\dot f_{(1)}t^2-\phi_{(1)}\text{,}
  \end{align}
\end{subequations}
with $\phi_{(1)}=-3L_0+\varpi_0$. The frequency and the time frequency shift are defined by
\begin{subequations}\label{eq:fdotf1}
\begin{align}
  f_{(1)}&=\frac{3n_0}{2\pi}\bigg(1+\frac{3\dot L_{1\mathrm{PN}}-\dot\varpi_{1\mathrm{PN}}}{3n_0}+\frac{3\dot L_{\mathrm{M}}-\dot\varpi_{\mathrm{M}}}{3n_0}\bigg)\text{,}\label{eq:f1}\\
  \dot f_{(1)}&=\frac{9n_0}{4\pi}\frac{|\dot a_{2.5\mathrm{PN}}|}{a_0}\text{,}
\end{align}
\end{subequations}
respectively. The ellipsis in the expressions \eqref{eq:h1ecc} correspond to the other first-order terms in eccentricity which occur at frequencies $L-\varpi$ and $L+\varpi$; we neglect them for the current discussion since their amplitudes are at least 3 and 4.5 times smaller than the dominant one occurring at frequency $3L-\varpi$, respectively. Hereafter, we refer to the GW signals $\{h_{+}^{(0)},h_{\times}^{(0)}\}$ and $\{h_{+}^{(1)},h_{\times}^{(1)}\}$ as the main signal (or the first harmonic signal) and the second harmonic signal, respectively.

The amplitude of the first and second harmonics (i.e., $\mathcal{A}_{(0)}$ and $\mathcal{A}_{(1)}$, respectively) can be further simplified when $t\ll t_{\mathrm{GW}}$. At zeroth-order in $t/t_{\mathrm{GW}}$, we have expressions as follows: $\mathcal{A}_{(0)}\simeq -h_0(1-5e_0{}^2/2)$ and $\mathcal{A}_{(1)}\simeq -9e_0h_0/4$. These are not accurate if $\sigma_{\mathcal{A}_{(i)}}$, the precision of the measured GW strain amplitudes, is the same order of magnitude than $t_{\mathrm{LISA}}/t_{\mathrm{GW}}$, where $i=\{0,1\}$ and $t_{\mathrm{LISA}}$ is the LISA mission duration. For GBs in the high frequency band of LISA, the characteristic time for the GWs radiation is $t_{\mathrm{GW}}\sim 10^4\ \mathrm{yr}$, that is to say three orders of magnitude larger than a 5-years LISA mission (i.e., $t_{\mathrm{LISA}}=5\ \mathrm{yr}$). Hence, the simplifications are effective as long as the precisions on the determination of the strain amplitudes satisfie: $\sigma_{\mathcal{A}_{(i)}}/\mathcal{A}_{(i)} \gg 10^{-3}$ with $i=\{0,1\}$. For GB sources, this condition translates into a constraint on the SNR of the source, namely $\mathrm{SNR}\ll 1000$ (see appendix \ref{sec:appB}).

Let us emphasize that we have kept a second order term in eccentricity in the expression of $\mathcal{A}_{(0)}$ [see Eq.~\eqref{eq:Amp0}]. The reason is linked to the fact that this is the only second order term in $e_0$ that is actually visible when processing the GW signal for the realistic case we are considering hereafter in Sect.~\ref{sec:result}.

According to Eqs. \eqref{eq:f0}, \eqref{eq:f1}, and \eqref{eq:Lmag}, one may infer that magnetism is responsible of shifting the frequencies in the GW signal with respect to the case without magnetism. Indeed, the frequency shifts are given by
\begin{subequations}
\begin{align}
  2\pi\Big(f_{(0)}-f_{(0)}|_{\bar U_{\mathrm{M}}=0}\Big)=4\dot\varpi_{\mathrm{M}}+\mathcal{O}(e_0{}^2),\\
  2\pi\Big(f_{(1)}-f_{(1)}|_{\bar U_{\mathrm{M}}=0}\Big)=5\dot\varpi_{\mathrm{M}}+\mathcal{O}(e_0{}^2),   
\end{align}
\end{subequations}
where $f_{(0)}|_{\bar U_{\mathrm{M}}=0}$ and $f_{(1)}|_{\bar U_{\mathrm{M}}=0}$ are respectively the frequency of the first and second harmonics when the magnetic energy is null. The shift is thus directly proportional to the magnetic contribution in the secular precession of the longitude of the pericenter. Moreover, as it may be seen from Eq. \eqref{eq:omegamag}, the precession of the longitude of the pericenter is directly proportional to the magnetic energy of the binary system, and hence, the frequency shift is thus proportional to the magnetic energy. Therefore, measuring the magnetic frequency shift is a way to estimate the secular magnetic energy within the source of the GWs.

To this end, one may actually notice a useful linear combination between the frequencies of the first and second harmonics, namely
\begin{equation}
  \dot\varpi_{\mathrm{M}}=3\pi f_{(0)}-2\pi f_{(1)}-\dot\varpi_{1\mathrm{PN}}.
  \label{eq:varpiM}
\end{equation}
This linear combination suppresses the main Keplerian contribution $n_0$ and returns the total precession of the longitude of the pericenter. Then, the modeling of the GR contribution (represented here by the dominant 1PN order) allows us to estimate the magnetic contribution and hence the magnetic energy. All this can be achieved at the condition that the second harmonic frequency can be measured, which necessitates that the eccentricity be high enough to render the second harmonic visible.

\begin{figure}
    \centering
    \includegraphics[scale=0.41]{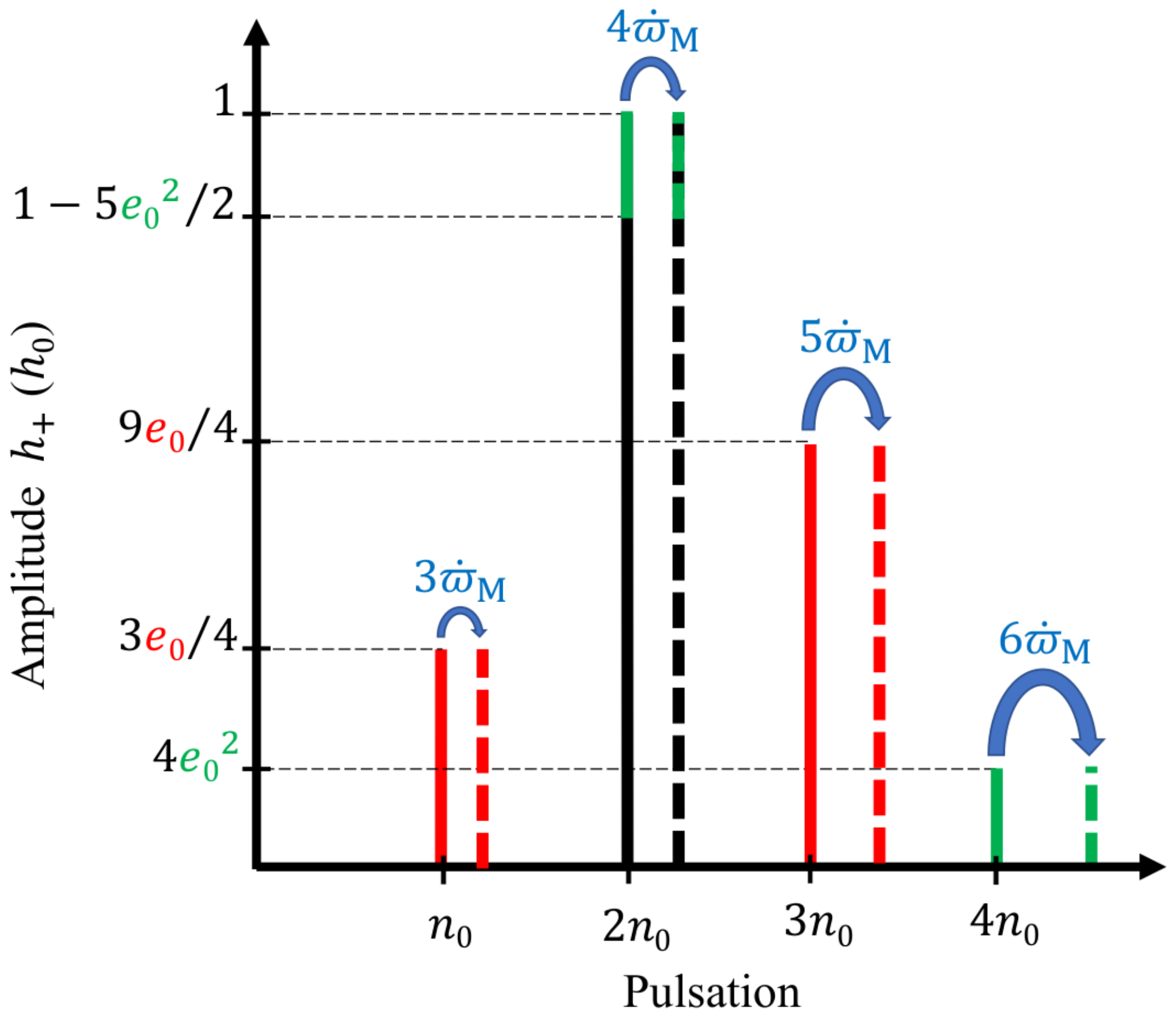}
    \caption{Schematic representation of the expected frequency spectrum of the ``+'' GW polarization for a magnetic GB in quasi-circular orbit up to second-order in eccentricity. The spectrum for the ``$\times$'' polarization is similar than the one for the $+$ polarization. The zeorth, first, and second-order contributions in eccentricity to the GW signal are depicted in \emph{black}, \emph{red}, and \emph{green}, respectively. The \emph{plain} and \emph{dashed} signals represent the frequency spectrum for a non-magnetic and a magnetic GB, respectively. Hence, a frequency peak occurring at pulsation $kn_0$ with $k\in\mathbb{N}^+$ is shifted by $(k+2)\dot{\varpi}_{\mathrm{M}}$ with respect to a configuration where magnetism is neglected.}
    \label{fig:freq_shift}
\end{figure}

In order to summarize the discussion, we represent in Fig.~\ref{fig:freq_shift} the frequency spectrum that is expected for a magnetic GB in quasi-circular orbit. The different harmonics are generated when the eccentricity is not null. At $\ell$th-order in eccentricity, we expect harmonics at pulsation $k_jn_0$ (considering only positive values) with $k_j=\ell+2(2-j)$, where $j=\{1,\ldots,(\ell+2)/2\}$ for $\ell$ even or $j=\{1,\ldots,(\ell+1)/2+1\}$ for $\ell$ odd. As seen from Eqs.~\eqref{eq:f0} and \eqref{eq:f1}, the harmonics are not expected exactly at pulsation $k_jn_0$ even when not considering magnetic effects; there are still the GR perturbation terms remaining. However the GR deviations are small before the Keplerian term $n_0$, so they are not represented in Fig.~\ref{fig:freq_shift}, which aims at emphasize the effects of eccentricity and magnetism only. Beside the first harmonic, we see from Fig. \ref{fig:freq_shift}, that the amplitude of the higher harmonics are at least linearly proportional to the eccentricity, meaning that the contribution of higher harmonics to the total GW signal is expected to decrease rapidly when $e\ll 1$. At $\ell$th-order in eccentricity, the amplitude of the harmonics are expected to scale as $\propto e^\ell h_0$. As discussed previously, and represented in Fig.~\ref{fig:freq_shift}, magnetism through the dipole-dipole interaction will shift all the frequencies of the GW signal with respect to a similar configuration where magnetism is neglected. The importance of the magnetic shift depends on the fundamental pulsation. A pulsation $kn_0$ with $k\in\mathbb{N}^+$ is thus expected to be shifted by the amount $(k+2)\dot\varpi_{\mathrm{M}}$ with respect to the non-magnetic case.

\subsection{Physical parameters from GW signal's parameters}
\label{sec:physfromGW}

When considering a binary system in circular orbit (i.e., $e_0=0$), the only remaining pieces in the expressions of the mode polarizations are the zeroth-order terms in eccentricity [cf. Eqs.~\eqref{eq:hpc1}]. Therefore, the data analysis can return measurements of the following set of the GW signal's parameters: $(\mathcal{A}_{(0)},\iota,f_{(0)},\dot{f}_{(0)})$. However, if the eccentricity of the binary system is sufficiently high so that the second harmonic has a SNR sufficient to be detected, then $\mathcal{A}_{(1)}$, $f_{(1)}$, and $\dot{f}_{(1)}$ can be determined too. Hence, the complete set of GW signal's parameters from a single source is actually given by $(\mathcal{A}_{(0)},\mathcal{A}_{(1)},\iota,f_{(0)},f_{(1)},\dot{f}_{(0)},\dot{f}_{(1)})$.

From the wave's parameters, we eventually aim at determining the physical parameters of the source, namely $(\eta,e_0,n_0,\mathcal{M},D,\iota,\bar U_{\mathrm{M}})$; from these, we can then determine the total mass $m$ from the definition of the chirp mass [cf. Eq.~\eqref{eq:chirp}] and the semi-major axis $a_0$ can be deduced from Kepler third law of motion [cf. Eq.~\eqref{eq:kep3}].

When considering GR at the 2.5PN approximation together with the magnetic dipole-dipole interaction, the time frequency shifts $\dot{f}_{(0)}$ and $\dot{f}_{(1)}$ are in fact linearly dependent, and this reduces the number of independent GW signal's parameters from 7 to 6. However, the number of physical parameters of the source is still 7, therefore, the system of equations is actually under-determined. Our best chance to make improvements is thus to assume a given value for the symmetric mass ratio $\eta$, and then find all the other physical parameters for the range of all possible values of $\eta$. For a binary system of similar masses, namely $m_2=m_1=m/2$, we have $\eta=1/4$ and $\mathcal{M}=0.25^{3/5}m\simeq0.435\,m$. On the other hand, for a binary system with $m_1\gg m_2$, the symmetric mass ratio reduces to $\eta\simeq m_2/m_1$ and $\mathcal{M}\simeq m_1(m_2/m_1)^{3/5}$ at first order in $m_2/m_1$. Therefore, the range of variation of $\eta$ is $0<\eta\leqslant 1/4$. Considering that compact GBs are expected to have similar masses, it is more likely that $\eta$ will remain close to $1/4$ for the sources we are considering in this study.

Considering the symmetric mass ratio as an input parameter, we now have to solve a system of non-linear equations to determine the set of parameters $(e_0,n_0,\mathcal{M},D,\iota,\bar U_{\mathrm{M}})$ from relations \eqref{eq:amp}, \eqref{eq:Amp0}, \eqref{eq:fdotf0}, \eqref{eq:Amp1}, and \eqref{eq:f1}. This transformation must be performed numerically for a better determination that takes into account correlations between the GW signal's parameters. However, first-order approximate solutions provide instructive simple relationships that turn out to provide a good precision in the context of inspiraling GBs in quasi-circular orbit. The approximate solutions are as follows:
\begin{subequations}
\begin{align}
  e_0&\simeq\frac{4}{9}\frac{\mathcal{A}_{(1)}}{\mathcal{A}_{(0)}}-\frac{160}{729}\left(\frac{\mathcal{A}_{(1)}}{\mathcal{A}_{(0)}}\right)^3\text{,}\label{eq:e0}\\
  n_0&\simeq\pi f_{(0)}-\mathcal{H}(e_0)\left[\frac{5\pi^{3/2}}{96\eta}f_{(0)}^{1/2}\dot f_{(0)}\frac{(1-e_0{}^2)}{\mathcal{G}(e_0)}\right]^{2/5}\text{,}\\
  \mathcal{M}&\simeq\frac{c^3}{G} \bigg[\frac{5}{96\pi^{8/3}}\frac{\dot{f}_{(0)}}{f_{(0)}^{11/3}}\frac{(1-e_0{}^2)^{7/2}}{\mathcal{G}(e_0)}\bigg]^{3/5}\text{,}\\
  D&\simeq\frac{5c}{48\pi^2}\frac{\dot f_{(0)}}{\mathcal{A}_{(0)}f_{(0)}^3}\frac{(1-\tfrac{5}{2}e_0{}^2)(1-e_0{}^2)^{7/2}}{\mathcal{G}(e_0)}\text{,}\\
  \bar U_{\mathrm{M}}&\simeq\frac{5}{288\pi^3}\frac{c^5}{G}\frac{\dot{\varpi}_{\mathrm{M}}\dot f_{(0)}}{f_{(0)}^{4}}\frac{(1-e_0{}^2)^{4}}{\mathcal{G}(e_0)}\text{,}\label{eq:mu1mu2}
    \end{align}
\end{subequations}
where the rate of precession of the longitude of the pericenter caused by the magnetic dipole-dipole interaction in given in term of $f_{(0)}$, $f_{(1)}$, and the GR perturbation in Eq. \eqref{eq:varpiM}, where $\dot\varpi_{1\mathrm{PN}}$ is given by
\begin{align}
  \dot\varpi_{1\mathrm{PN}}&\simeq\left[\frac{15\sqrt{3}\,\pi^{3/2}}{32\eta}f_{(0)}^{1/2}\dot{f}_{(0)}\frac{(1-e_0{}^2)}{\mathcal{G}(e_0)}\right]^{2/5}\text{.}
\label{eq:varpiGR}
\end{align}

Let us emphasize that the detection of eccentricity can be achieved once $\mathcal{A}_{(1)}$, the amplitude of the second harmonic, can be measured as shown in Eqs. \eqref{eq:e0}. Similarly, because the eccentricity is responsible for generating the harmonic at frequency $f_{(1)}$, we see from Eq.~\eqref{eq:mu1mu2} and \eqref{eq:varpiM} that the detection of a magnetic dipole-dipole interaction within the binary system cannot be done for a circular orbit. Indeed, a circular motion induces a unique quasi-monochromatic GW radiation at frequency $f_{(0)}$ meaning that $\dot{\varpi}_{\mathrm{M}}$ cannot be measured from GWs radiation alone as anticipated in \citet{PhysRevD.105.124042}.

\section{Simulation of a realistic gravitational waveform}
\label{sec:numerical_setup}

In this section, we use Eqs. \eqref{eq:hphc} in order to generate, in the time domain, the GW signal emitted by a magnetic GB in quasi-circular orbit. Then, we generate the different Time Delay Interferometry (TDI) channels characterizing the LISA response to the incoming GW signal in the frequency domain with the help of the LDC tools~\cite{LDC}. We apply standard statistical analysis techniques in order to infer the physical parameters of the simulated GB in the next section.

\subsection{HM Cancri (RX J0806.3+1527)}

Let us consider the  verification binary HM Cancri (RX J0806.3+1527) whose two massive components are orbiting each other at a frequency of $6\ \mathrm{mHz}$. Thus, the GW signal is expected in the LISA low frequency band. According to \citet{2020MNRAS.491.3000M}, the X-ray emission and the optical light-curve modulations of HM Cancri are both compatible with an eccentricity at the level of 0.1, making it one of the most promising sources for detecting multiple harmonics. In addition, HM Cancri is composed of two WDs whose surface magnetic fields could be as high as $10^{9}\ \mathrm{G}$ \cite{2020AdSpR..66.1025F}. Therefore, if the frequency of the first harmonic can be measured by LISA, HM Cancri is a good candidate for testing whether or not the strength of the magnetic dipole-dipole interaction can be extracted from the GW data alone. However, it seems that no information is presently available on the nature and intensity of magnetic fields in HM Cancri. In order to investigate if magnetism could be determined from LISA data, we assume the most favorable case where the two WDs have magnetic moments at the level of $10^{33}\ \mathrm{A}\,\mathrm{m}^2$, which corresponds to magnetic fields at the level of $10^{9}\ \mathrm{G}$. In addition, we assume that the magnetic moments are exactly in the most secularly stable configuration which corresponds to the minimum of the secular magnetic energy [cf. Eq. \eqref{eq:magE}]. This is given when the magnetic moments are both orthogonal to the orbital plane and are in opposite direction (Aykroyd \emph{et al}. in prep.), namely $\mu_1^z=\mu_1=\Vert\bm\mu_1\Vert$, $\mu_2^z=-\mu_2=-\Vert\bm\mu_2\Vert$, and $\bm\mu_1\cdot\bm\mu_2=-\mu_1\mu_2$. Accordingly, the secular magnetic energy reduces to
\begin{equation}
  \bar{U}_{\mathrm{M}}=-\frac{\mu_0\mu_1\mu_2}{2\pi a_0{}^3(1-e_0{}^2)^{3/2}}.
\end{equation}

All the physical parameters that we use for simulating GW emission from HM Cancri are recalled in Tab. \ref{tab:GB_param}.

\begin{table}
    \centering
    \caption{HM Cancri parameters. Some of the GW signal's parameters ($f_0,\dot f_0,\iota,\lambda,\beta$) are deduced from electromagnetic observations. The others are either deduced from simplified relationship (e.g., $\mathcal{A}_0$) or chosen arbitrarily (e.g., $\phi$ and $\psi$).}
    \begin{tabularx}{\linewidth}{cc|>{\centering\arraybackslash}Xc}
    \hline\hline
    $\Big.$Parameter & Unit & Value & Refs.\\
    \hline
    \multicolumn{4}{c}{$\Big.$GW signal's parameter}\\
    \hline
    $\mathcal{A}_0$ & - &  $-2.8 \times 10^{-22}$ & \cite{Kupfer_2023}\\
    $\iota$ & rad & $0.663225$ & \cite{Strohmayer_2004,Roelofs_2010,Kupfer_2018,Kupfer_2023}\\ 
    $f_0$ & $\mathrm{mHz}$  & $6.220279$ & \cite{Strohmayer_2004,Roelofs_2010,Kupfer_2018,Kupfer_2023}\\
    $\dot{f}_0$ & $\mathrm{Hz}^2$  & $ 3.6 \times 10^{-16}$ & \cite{Strohmayer_2004,Roelofs_2010,Kupfer_2018,Kupfer_2023}\\
    $\phi$ & $\mathrm{rad}$ & $1.570796$ & chosen\\
    $\psi$ & $\mathrm{rad}$  & $1.570796$ & chosen\\
    $\lambda$ & $\mathrm{rad}$ & $0.082$ & \cite{GAIA_edr3}\\ 
    $\beta$ & $\mathrm{rad}$ & $2.102$ & \cite{GAIA_edr3}\\ 
    \hline
    \multicolumn{4}{c}{$\Big.$Physical parameter}\\
    \hline
    $e_0$ & -  &  $0.1$ & \cite{2020MNRAS.491.3000M}\\
    $m_1$ & $\mathrm{M}_\odot$ & $0.55$ & \cite{Strohmayer_2004,Roelofs_2010,Kupfer_2018,Kupfer_2023}\\
    $m_2$ & $\mathrm{M}_\odot$ & $0.27$ & \cite{Strohmayer_2004,Roelofs_2010,Kupfer_2018,Kupfer_2023}\\
    $D$ & $\mathrm{kpc}$ & 7.5 & \cite{Strohmayer_2004,Roelofs_2010,Kupfer_2018,Kupfer_2023}\\
    $n_0$ & $\mathrm{rad}\,\mathrm{s}^{-1}$ & $2.083583\times10^{-3}$ & $\pi f_0$\\
    $a_0$ & $\mathrm{km}$ & $292\,692$ & Eq. \eqref{eq:kep3}\\
    $\eta$ & - & 0.22 & Eq. \eqref{eq:chirp}\\
    $\mathcal{M}$ & $\mathrm{M}_\odot$ & 0.33 & Eq. \eqref{eq:chirp}\\
    $\mu_1$ & $\mathrm{A}\,\mathrm{m}^{-2}$ & $10^{33}$ & chosen\\
    $\mu_2$ & $\mathrm{A}\,\mathrm{m}^{-2}$ & $10^{33}$ & chosen\\   
    $\bar U_{\mathrm{M}}$ & J & $-1.41\times 10^{36}$ & Eq. \eqref{eq:magE}\\   
    \hline
    \end{tabularx}
    \label{tab:GB_param}
\end{table}

HM Cancri being a verification binary, its orbital frequency and time frequency shift are already known from electromagnetic observations \cite{Strohmayer_2004}. Therefore, assuming a circular motion, the expected GW main frequency $f_0$ and expected time frequency shift $\dot f_0$ can be determined too. In addition, by omitting correction terms due to GR and magnetism in Eqs. \eqref{eq:fdotf0}, the mean motion then reduces to $n_0\simeq\pi f_0$ and hence the expected amplitude of the GW signal [see Eq.~\eqref{eq:amp}] may be defined as
\begin{equation}
  \mathcal{A}_0\equiv-\frac{2c\,(\pi f_0)^{2/3}}{D}\left(\frac{G\mathcal M}{c^3}\right)^{5/3}\text{,}
  \label{eq:h0simp}
\end{equation}
which is equivalent to expression (4) of \citet{Kupfer_2023}. Substituting the values for the masses and observed frequency in the previous equation, we end up with the estimated value of $\mathcal{A}_0$ shown in Tab. \ref{tab:GB_param}.

Beside $f_0$ and $\dot f_0$, there are other parameters that can be inferred from electromagnetic observations of HM Cancri, for instance the inclination $\iota$ and the sky localisation (i.e., the ecliptic latitude $\lambda$ and the ecliptic longitude $\beta$). Two additional parameters are also reported in Tab.~\ref{tab:GB_param}: $\phi$ and $\psi$; they represent the phase at the origin and the polarization angle, respectively. We arbitrarily set their numerical values to $\pi/2$.

\subsection{LDC waveform simulations}

To represent GWs from eccentric and magnetic GBs, we will use the framework provided by the LISA data challenge LDC \cite{LDC}. The framework allows us to generate all three TDI data channels, $A,E,T$ \cite{TDI1999,TDI2002} based on LISA's analytical orbits and on the mode polarizations $\{+,\times\}$ described in the Sect. \ref{sec:waveform}. In our case, the simplest approach is to generate the waveform modeling directly in the time domain. This is done after substituting the secular solutions \eqref{eq:solpiL} and \eqref{eq:solaz} into Eqs. \eqref{eq:hphc}. This allows us to simulate HM Cancri's GW signal in the time domain considering GR, magnetism, and eccentricity up to second order. We recall that the input numerical values of the physical parameters are reported in Tab. \ref{tab:GB_param}.

Then, to stress the effects of magnetism and eccentricity of the binary system on the gravitational waveform, we simulate two different physical cases :
\begin{enumerate}
    \item NGB : a non-magnetic GB in circular orbit (i.e., $\mu_1=\mu_2=0$, $e_0=0$),
    \item EMGB : a magnetic GB in quasi-circular orbit (i.e., $\mu_1$, $\mu_2$, and $e_0$ fixed to their values in Tab. \ref{tab:GB_param}).
\end{enumerate}

In order to report efficiently the expected EMGB's frequency shifts of the first and second harmonic with respect to the main frequency of the NGB case, we introduce the parameter $\Delta f_{(i)}$, being defined such as
\begin{equation}
  \Delta f_{(i)}=f_{(i)}-\frac{(i+2)}{2}f_0 \qquad \mathrm{with} \qquad i=\{0,1\}.
  \label{eq:expDeltaf}
\end{equation}
We call $\Delta f_{(i)}$ the expected relative frequency shift. We recall that $f_{(i)}$ with $i=\{0,1\}$ are the analytical expressions of the first and second harmonic frequencies [see Eqs.~\eqref{eq:fdotf0} and \eqref{eq:fdotf1}] while $f_0$ is the expected value of the frequency of the gravitational radiation when considering a binary system in circular orbit and without magnetic effects. From this definition, we should expect
\begin{equation}
  f_0\equiv\frac{n_0}{\pi}\left[1+\eta^{3/5}n_0{}^{2/3}\left(\frac{G\mathcal{M}}{c^3}\right)^{2/3}\right].
  \label{eq:expf0}
\end{equation}
In Tab. \ref{tab:GB_param}, we gave the numerical value of $f_0$ when considering that it is given from the orbital period which is itself inferred from electromagnetic observations.

Let us also introduce the parameter $\gamma_{(i)}$ representing the ratio between the analytical expressions of the first and second harmonic amplitudes [see Eqs. \eqref{eq:Amp0} and \eqref{eq:Amp1}] at the initial time $t_0$ and the amplitude $\mathcal{A}_0$ [cf. Eq. \eqref{eq:h0simp}], that is to say
\begin{equation}
  \gamma_{(i)}=\left.\frac{\mathcal{A}_{(i)}}{\mathcal{A}_0}\right|_{t_0} \qquad \mathrm{with} \qquad i=\{0,1\}.
  \label{eq:expgamma}
\end{equation}
We call $\gamma_{(i)}$ the expected amplitude ratio.

\begin{table}
	\centering
	\caption{Numerical values of the expected relative frequency shifts $\Delta f_{(i)} $, the time frequency shifts $\dot{f}_{(i)}$, and the relative amplitudes ratio $\gamma_{(i)}$ with $i=\{0,1\}$ for the NGB and EMGB cases.}
	\begin{tabularx}{\linewidth}{c|>{\centering\arraybackslash}X>{\centering\arraybackslash}X>{\centering\arraybackslash}X}
	\hline\hline
	$\Big.$Case    &  $\Delta f_{(i)}$ [nHz] & $\dot{f}_{(i)}$ [$\mathrm{nHz}\,\mathrm{yr}^{-1}$] & $\gamma_{(i)}$ [ - ]\\
	\hline
	\multicolumn{4}{c}{$\Big.$ First harmonic ($i = 0$)} \\
	\hline
	NGB     & $ 0.0$ & $23.62$ & $1.0$\\
	EMGB    & $ -92.29$ & $25.21$ & $0.975$\\
	\hline
	\multicolumn{4}{c}{$\Big.$ Second harmonic ($i = 1$)} \\
	\hline
	NGB     & -    & - & -\\
	EMGB    & $ -117.73$ & $37.82$ & $0.225$\\
	\hline
	\end{tabularx}
	\label{tab:EMGB_f0fdotA}
\end{table}

In Tab. \ref{tab:EMGB_f0fdotA}, we give the numerical values of $\Delta f_{(i)}$ and $\gamma_{(i)}$ for $i=\{0,1\}$ according to the input values of the physical parameters presented in Tab. \ref{tab:GB_param}. Since NGB is expected at the main frequency only, it does not contribute to the second harmonic. In addition, as it may be anticipated from the definitions in Eqs. \eqref{eq:expDeltaf} and \eqref{eq:expf0}, the expected frequency shift of the main frequency of the NGB case (i.e., $\Delta f_{(0)}$) is in fact null. For the second harmonic, the frequency shift of the EMGB case is not null as one could expect. Let us recall, that in this case the shift is measured from $3f_0/2$, namely from a frequency remaining close to the second harmonic frequency. As a matter of fact, $\Delta f_{(1)}$ is proportional to 1PN and magnetic terms according to the following relationship
\begin{equation}
  2\pi\Delta f_{(1)}=3\big(\dot{L}_{1\mathrm{PN}}-\dot{L}_{1\mathrm{PN}}|_{e=0}+\dot{L}_{\mathrm{M}}\big)-\dot{\varpi}_{1\mathrm{PN}}-\dot{\varpi}_{\mathrm{M}}.
\end{equation}

Concerning the expected amplitude ratio, we see from Tab. \ref{tab:EMGB_f0fdotA} that the amplitude of the main signal is decreased when passing from the NGB to the EMGB case. Indeed, $\gamma_{(0)}$ is mainly departing from 1 because of terms that are of second order in eccentricity:
\begin{equation}
  \gamma_{(0)}=1-\frac{5}{2}e_0{}^2.
\end{equation}
While computing this ratio, we omitted the contributions from 1PN and magnetism terms that are completely negligible before eccentricity. The expected amplitude ratio for the second harmonic is simply given by its dominant contribution coming from eccentricity for similar reasons, therefore we have the following expressions:
\begin{equation}
  \gamma_{(1)}=\frac{9}{4}e_0.
\end{equation}

Numerical estimates presented in Tab. \ref{tab:EMGB_f0fdotA} allows us to confirm that for the case of HM Cancri the separation between the main frequency and the second harmonic's frequency is much larger (of the order of the mHz) that the relative shifts induced by the physical effects (of the order of tenth of nHz). In addition, the time frequency shifts of the first and second harmonics (i.e., $\dot f_{(0)}$ and $\dot f_{(1)}$) induce frequency changes at the level of tenth of nHz after few years of observation. These two conditions ensure that each harmonic of the GW signal in Eq. \eqref{eq:hphc} do not overlap and can in fact be studied independently.

The inference of the physical parameters of HM Cancri from the independent study of each harmonic in the GW signal is the subject of the next section.

\section{Analysis of the simulated waveform}
\label{sec:result}

The LDC tools provide a GW simulator for GBs that depends on eight parameters representing a quasi-monochromatic GW signal: ($\mathcal A,\iota,f,\dot f,\lambda,\beta,\psi,\phi$). Among these, the ecliptic latitude $\lambda$, the ecliptic longitude $\beta$, the inclination $\iota$, and the polarization $\psi$ are the extrinsic parameters describing the position and orientation of the GWs' source with respect to LISA's frame. The remaining four parameters, namely the amplitude $\mathcal{A}$, the frequency $f$, the frequency shift $\dot f$, and the initial phase $\phi$ are all intrinsic parameters; their value are computed at an initial instant of time $t_0$. Intrinsic and extrinsic parameters dictate the temporal evolution of $h_{+}^{\mathrm{LDC}}$ and $h_{\times}^{\mathrm{LDC}}$, namely the $\{+,\times\}$ mode polarizations of a quasi-monochromatic GW signal occuring at frequency $f$ as modelled within the LDC:
\begin{subequations}\label{eq:GB_WF}
  \begin{align}
    h_{+}^{\mathrm{LDC}}(t) &= \mathcal{A}\left(1+\cos^2\iota\right) \cos \Phi(t), \\
    h_{\times}^{\mathrm{LDC}}(t) &= 2 \mathcal{A} \cos\iota \sin \Phi(t),
  \end{align}
\end{subequations}
with
\begin{equation}
  \Phi(t) = 2\pi f t + \pi \dot{f} t^2 - \phi .
\end{equation}

According to the fact that the harmonics of the GW signal in Eq. \eqref{eq:hphc} do not overlap (see discussion at the end of Sect. \ref{sec:numerical_setup}), each harmonic of the signal can be analyzed independently from each other in the frequency domain using the LDC template \eqref{eq:GB_WF}. The LDC algorithm allows us to infer the GW signal's parameters by matching the quasi-monochromatic template \eqref{eq:GB_WF} to the simulated data is called ``fastGB''.

\subsection{Numerical setup}

The GW signal from an eccentric and a magnetic source is simulated by taking into account all the orders in the expansion described in Eq. \eqref{eq:hphc} and constitutes the data set on which the analysis algorithm works. As discussed in the previous section, the effective independence of the different harmonics allows to search for them using the LDC template \cite{LDC}. All results presented in this section are obtained with a Monte Carlo Markov Chain (MCMC) sampler with $10^{6}$ samples minus a 25\% burn-in. 
Two types of data analysis are conducted in this paper :
\begin{enumerate}
    \item ``Optimistic'': This case corresponds to an ideal search assuming a perfect knowledge of the inclination (see Fig. \ref{fig:orbit} in the detector adapted frame) and the sky localisation. This case is slightly optimisticic for HM Cancri since the inclination is currently poorly known. However, the 4th edition of the Gaia catalogue \cite{Gaia} is expected for a release at the same time than the LISA mission. Thus, it could provide a better determination of the inclination with enough precision to consider that these parameters are known in the analysis.
    \item ``Pessimistic'': This second case corresponds to an agnostic search for the GW signal, namely we provide no \emph{a priori} information to the sampler. For HM Cancri, this assumption is slightly pessimistic, for instance the position in the sky is known thanks to the Gaia mission with actually a better precision than LISA's. However, this approach is agnostic and could therefore be consistent with the results that would be expected if a magnetic and eccentric GB was not identified in advance. This approach comes close to the numerous methods proposed in the LDC for the detection of the ten thousand GBs expected over the duration of the mission.
\end{enumerate}

\subsection{Sampler results}

The following plots (cf. Figs. \ref{fig:Optimistic_1yr_main}--\ref{fig:Both_8yr_secondary}) will be presented as corner plots of the sampler results for each binary type (i.e., NGB and EMGB), for each duration (1-year, 4-years, and 8-years) and for the first and second harmonic peaks only. Indeed, the different harmonics that are generated by the second order terms in eccentricity cannot be observed for HM Cancri's eccentricity, except the signal at frequency $2L$ which modifies the amplitude of the first harmonic [see Eq. \eqref{eq:Amp0}]. In addition, among the harmonics generated at first order in eccentricity, only the harmonic at frequency $3L-\varpi$ can fully be identified by the algorithm. The two types of binaries are studied independently but represented on the same graphs for the sake of readability.
The figures represent the marginal 2D (contours) and 1D posterior distributions of the Bayesian analysis. The 1D posterior distribution of each parameter is on the diagonal while the correlation between two parameters is represented by the off-diagonal plots, namely the marginal 2D distributions.

\subsubsection{One-year}

\begin{figure}
    \includegraphics[width=0.48\textwidth]{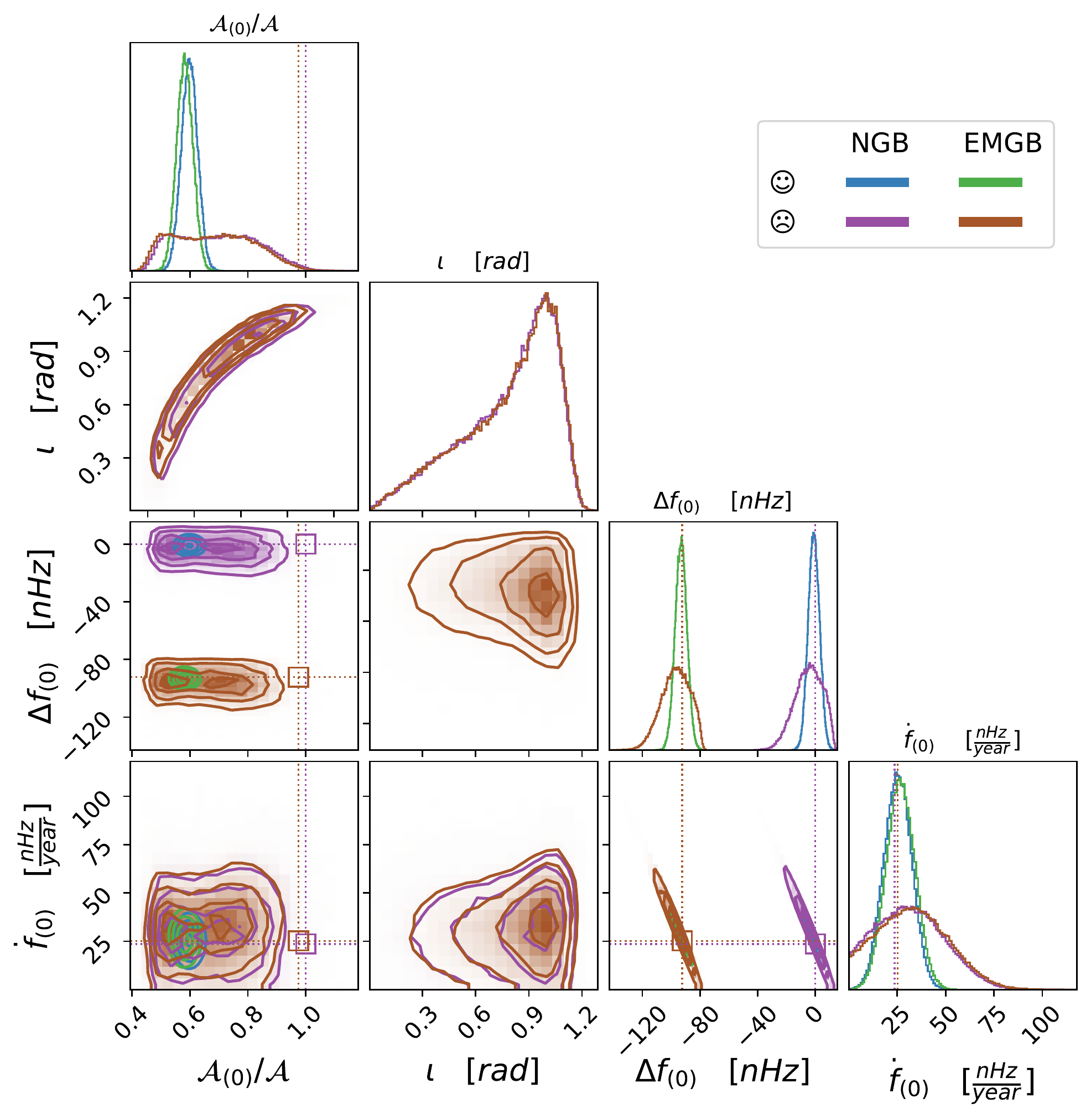}
    \caption{Optimistic (\smiley{}) and pessimistic (\frownie{}) studies for the main frequency after a 1-year LISA mission. Except for amplitude and inclination in the pessimistic case and the amplitude for the optimistic case, the parameters are well estimated as shown by the gaussian distributions. The dotted lines and the square marker represent the value injected during the simulation.}
    \label{fig:Optimistic_1yr_main}    
\end{figure}

For a 1-year LISA mission (see Fig. \ref{fig:Optimistic_1yr_main} and Tab. \ref{tab:EMGB_f0fdotA_1yr}), the uncertainty on the main frequency is sufficient to notice the frequency shift introduced by magnetism compared to non-magnetic GB in both the optimistic and pessimistic cases. However, it is important to note that the true Bayesian inference concerns the frequency $f_{(0)}$ and not directly the frequency shift, and hence no information concerning magnetism nor eccentricity can be deduced from the measure of $f_{(0)}$ alone. For this to happen, the frequency and amplitude of the second harmonic must be measured as well (see discussion in Sect. \ref{sec:phys_inter}). Unfortunately, for a 1-year LISA mission, the SNR of the second harmonic signal is too low to be detected for HM Cancri's eccentricity. Therefore, no information can be drawn on eccentricity nor magnetism.

\begin{table}
	\centering
	\caption{Numerical values of the measured relative frequency shift $\Delta f_{(0)}$, the time frequency shift $\dot{f}_{(0)}$, and relative amplitude ratio $\gamma_{(0)}$ for NGB and EMGB for the optimistic and pessimistic cases and a 1-year LISA mission simulation.}
	\begin{tabularx}{\linewidth}{c|>{\centering\arraybackslash}X c >{\centering\arraybackslash}X >{\centering\arraybackslash}X}
	\hline
	\hline
	$\Big.$1-year & $\Delta f_{(0)}$ & $\dot{f}_{(0)}$  & \multicolumn{2}{c}{$\gamma_{(0)}$}\\
	$\Big.$ & [nHz] & [$\mathrm{nHz}\,\mathrm{yr}^{-1}$] & \multicolumn{2}{c}{[ - ]}\\
	\hline
	NGB     & $ -5 \pm 4$ & $25 \pm 7$ & $0.6 \pm 0.03$ & $0.7 \pm 0.1$\\
	EMGB    & $ -97 \pm 4$ & $27 \pm 8$ & $0.55 \pm 0.03$ & $0.6 \pm 0.1$\\
	\hline
	$\Big.$Case & \multicolumn{2}{c}{Optimistic and pessimistic} & Optimistic & Pessimistic \\
	\hline
	\end{tabularx}
	\label{tab:EMGB_f0fdotA_1yr}
\end{table}

The mean value of the frequency (second column of Tab. \ref{tab:EMGB_f0fdotA_1yr}) is in agreement with the expected theoretical values (second column of Tab. \ref{tab:EMGB_f0fdotA}). The amplitude $\mathcal{A}_{(0)}$ and the inclination $\iota$ are highly correlated as expected in the waveform formulation (\ref{eq:hpc1}). This correlation takes the form of a bias in the estimation of the parameters. We will see hereafter that the bias decreases with the mission duration.

\subsubsection{Four-years}

\begin{figure}
    \centering
    \begin{subfigure}[b]{0.48\textwidth}
        \centering
        \includegraphics[width=\textwidth]{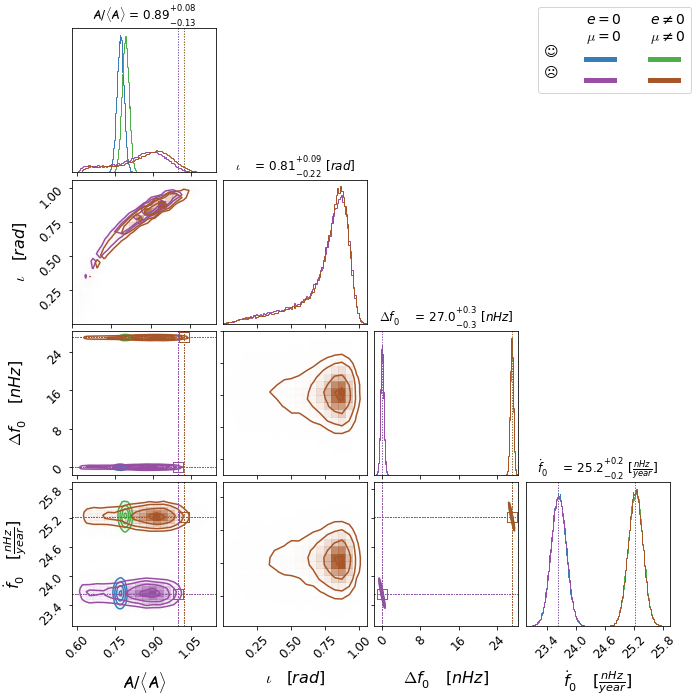}
        \caption{First harmonic}
        \label{fig:Both_4yr_main}
        \end{subfigure}%
    \\
    \begin{subfigure}[b]{0.48\textwidth}
        \centering
        \includegraphics[width=\textwidth]{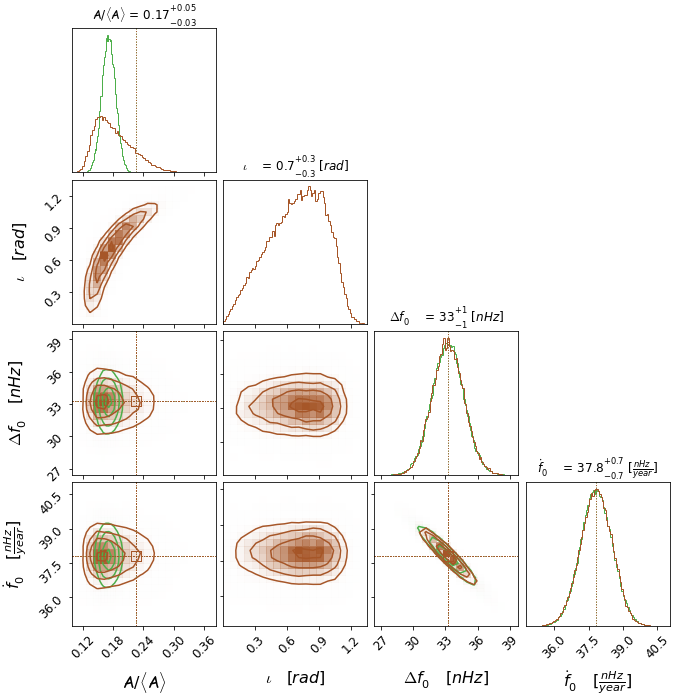}
        \caption{Second harmonic}
        \label{fig:Both_4yr_secondary}
    \end{subfigure}  
    \caption{Optimistic (\smiley{}) and pessimistic (\frownie{}) studies for both frequencies and a $4$-years mission. Except for amplitude and inclination in the pessimistic case and the amplitude for the optimistic case, the parameters are well estimated as shown by the gaussian distributions. The dotted lines and the square marker represent the value injected during the simulation.}
\end{figure}

For a 4-years LISA mission (see Fig. \ref{fig:Both_4yr_main} and Tab.~\ref{tab:EMGB_f0fdotA_4yr}), the uncertainty on the main frequency is still small enough to visually discriminate between the NGB and EMGB cases. In addition, the uncertainty on the time derivative of the main frequency is small enough to visually observe the influence of the eccentricity. The amplitude/inclination bias decreases with respect to the 1-year LISA mission scenario but is still present.

Regarding the second harmonic peak (see Fig. \ref{fig:Both_4yr_secondary}), its SNR is now high enough so that the LDC algorithm can actually detect it. The same amplitude/inclination bias than for the first harmonic is also present for the second harmonic. 

\begin{table}
	\centering
	\caption{Numerical values of the measured relative frequency shift $\Delta f_{(0)}$, the time frequency shift $\dot{f}_{(0)}$, and relative amplitude ratio $\gamma_{(0)}$ for NGB and EMGB for the optimistic and pessimistic cases and a 4-year LISA mission simulation.}
	\begin{tabularx}{\linewidth}{c|>{\centering\arraybackslash}X c >{\centering\arraybackslash}X >{\centering\arraybackslash}X}
	\hline
	\multicolumn{5}{c}{$\Big.$First harmonic ($i = 0$)} \\
	\hline
	$\Big.$4-year & $\Delta f_{(0)}$ & $\dot{f}_{(0)}$  & \multicolumn{2}{c}{$\gamma_{(0)}$}\\
	$\Big.$ & [nHz] & [$\mathrm{nHz}\,\mathrm{yr}^{-1}$] & \multicolumn{2}{c}{[ - ]}\\
	\hline
	NGB     & $ -0 \pm 0.3$ & $23.6 \pm 0.2$ & $0.77 \pm 0.01$ & $0.8 \pm 0.1$\\
	EMGB    & $ -92.3 \pm 0.3$ & $25.2 \pm 0.2$ & $0.72 \pm 0.01$ & $0.8 \pm 0.1$\\
	\hline
	\multicolumn{5}{c}{$\Big.$Second harmonic ($i = 1$)} \\
	\hline
	NGB   & - & - & - & -\\
	EMGB    & $ -118 \pm 1$ & $37.8 \pm 0.7$ & $0.17 \pm 0.01$ & $0.18 \pm 0.04$\\
	\hline
	$\Big.$Case & \multicolumn{2}{c}{Optimistic and pessimistic} & Optimistic & Pessimistic \\
	\hline
	\end{tabularx}
	\label{tab:EMGB_f0fdotA_4yr}
\end{table}

Taking advantage of the fact that the posterior frequency distributions are totally confounded in Figs. \ref{fig:Both_4yr_main} and \ref{fig:Both_4yr_secondary} for the optimistic and pessimistic cases after a 4-years LISA mission scenario, we do not display them for the following 8-years scenario.

\subsubsection{Eight-years}

\begin{figure*}
    \centering
    \begin{subfigure}[b]{0.5\textwidth}
        \centering
        \includegraphics[height=8cm]{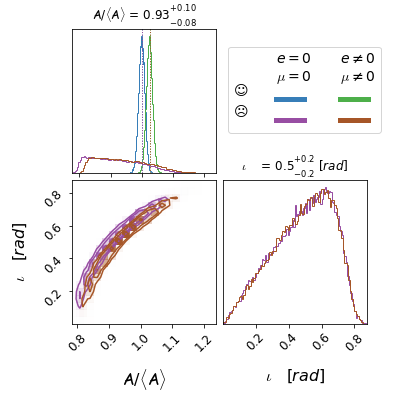}        
        \caption{Main signal.}
        \label{fig:Both_8yr_main}
        \end{subfigure}%
    \begin{subfigure}[b]{0.5\textwidth}
        \centering
        \includegraphics[height=8cm]{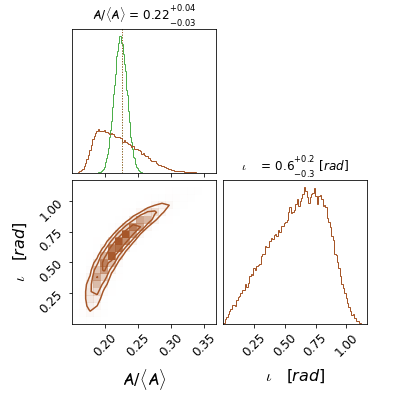}
        \caption{First harmonic}
        \label{fig:Both_8yr_secondary}
    \end{subfigure}  
    \caption{Optimistic (\smiley{}) and pessimistic (\frownie{}) studies for the both frequencies and a $8$-years mission.}
\end{figure*}

For a 8-years LISA mission (see Figs. \ref{fig:Both_8yr_main} and \ref{fig:Both_8yr_secondary} and Tab. \ref{tab:EMGB_f0fdotA_8yr}), the amplitude bias disappears and all the GW parameters are now well estimated. Let us emphasize that the estimated amplitude of the main frequency is actually higher for the NGB case than for the EMGB case. This reflects the contribution from second-order term in eccentricity as shown in Eq. \eqref{eq:Amp0}.

\begin{table}
	\centering
	\caption{Numerical values of the measured relative frequency shift $\Delta f_{(0)}$, the time frequency shift $\dot{f}_{(0)}$, and relative amplitude ratio $\gamma_{(0)}$ for NGB and EMGB for the optimistic and pessimistic cases and a 8-year LISA mission simulation.}
	\begin{tabularx}{\linewidth}{c|>{\centering\arraybackslash}X c >{\centering\arraybackslash}X >{\centering\arraybackslash}X}
	\hline
	\multicolumn{5}{c}{$\Big.$First harmonic ($i = 0$)} \\
	\hline
	$\Big.$8-year & $\Delta f_{(0)}$ & $\dot{f}_{(0)}$  & \multicolumn{2}{c}{$\gamma_{(0)}$}\\
	$\Big.$ & [nHz] & [$\mathrm{nHz}\,\mathrm{yr}^{-1}$] & \multicolumn{2}{c}{[ - ]}\\
	\hline
	NGB     & $ 0 \pm 0.09$ & $23.62 \pm 0.02$ & $1 \pm 0.01$ & $0.92 \pm 0.08$\\
	EMGB    & $ -92.3 \pm 0.09$ & $25.21 \pm 0.02$ & $0.93 \pm 0.01$ & $0.85 \pm 0.07$\\
	\hline
	\multicolumn{5}{c}{$\Big.$Second harmonic ($i = 1$)} \\
	\hline
	NGB   & - & - & - & -\\
	EMGB    & $ -84.1 \pm 0.4$ & $37.82 \pm 0.09$ & $0.22 \pm 0.01$ & $0.22 \pm 0.03$\\
	\hline
	$\Big.$Case & \multicolumn{2}{c}{Optimistic and pessimistic} & Optimistic & Pessimistic \\
	\hline
	\end{tabularx}
	\label{tab:EMGB_f0fdotA_8yr}
\end{table}

\subsection{Eccentricity and magnetism estimates}

In this subsection, we seek to estimate the value of the physical parameters from the GW parameters obtained in the previous subsection. We start with the eccentricity which depends on the amplitude of the main and second harmonics. The eccentricity is needed in order to estimate the rate of precessions of the longitude of the pericenter due to the magnetic dipole-dipole interaction.

\subsubsection{Eccentricity}
Considering a 4 or 8-years LISA mission, it is possible to estimate the value of the eccentricity from Eq. (\ref{eq:e0}) in both pessimistic and optimistic cases. By propagating the amplitude uncertainties on eccentricity, we find:
\begin{equation}
  \left(\frac{\sigma_{e_0}}{e_0}\right)^2 = \left(\frac{\sigma_{\mathcal{A}_{(0)}}}{\mathcal{A}_{(0)}}\right)^2 + \left(\frac{\sigma_{\mathcal{A}_{(1)}}}{\mathcal{A}_{(1)}}\right)^2.
  \label{eq:eccuncert}
\end{equation}
Neglecting the inclination/amplitude correlation, we show in appendix \ref{sec:appB} that the relative uncertainty in amplitude is in fact equivalent to the inverse of the SNR which depends essentially on the mission duration. Therefore, we expect the relative precision in eccentricity to vary such as $\sigma_{e_0}/{e_0} \propto $ 1/SNR. As discussed in Sect. \ref{sec:detectorframe}, for a mission duration less than 10 years and a SNR less than 1000, we can assume that the amplitude does not vary so that the eccentricity uncertainty is only limited by the amplitude uncertainty. The eccentricity estimate that is inferred from the data analysis is coherent with the injected eccentricity (cf. Tab. \ref{tab:GB_param}, $e_0=0.1$).

\begin{table}
	\centering
	\caption{HM Cancri eccentricity estimates for different LISA mission duration scenarios for both optimistic and pessimistic cases.}
	\begin{tabularx}{\linewidth}{c|>{\centering\arraybackslash}X >{\centering\arraybackslash}X}
	 \hline\hline
	 $\Big.$Case & Pessimistic & Optimistic \\
	 \hline
	4-years & $0.1 \pm 0.02$ & $0.101 \pm 0.008$ \\
	8-years & $0.11 \pm 0.02$ & $0.102 \pm 0.005$ \\
	\hline
	\end{tabularx}
	\label{tab:eccentricity}
\end{table}

Although the amplitude estimate is biased due to its correlation with the inclination, the bias is actually the same for both first and second harmonics [see Eqs. \eqref{eq:hpc1} and \eqref{eq:h1ecc}] so that the determination of the value of the eccentricity [see Eq.~\eqref{eq:e0}] is actually independent of it. However, according to Eq. \eqref{eq:eccuncert}, the uncertainty on the eccentricity depends on the uncertainties on the first and second harmonic's amplitude. These are in fact highly impacted by the inclination/amplitude bias. As a matter of fact, we can see in Tab. \ref{tab:eccentricity}, that the uncertainty barely changes between a 4 and a 8-years LISA mission scenario but is improved from $20\%$ to $5\%$ when passing from the pessimistic to optimistic case, respectively. 

\subsubsection{Magnetism}

From the eccentricity estimate, we can now determine the rate of precession of the longitude of the pericenter caused by the magnetic dipole-dipole interaction. In order to extract the frequency shift information due to magnetism, we first need to model the 1PN contribution in the total precession of the longitude of the pericenter, namely $\dot\varpi_{1\mathrm{PN}}$ [see Eq. \eqref{eq:varpiGR}]. Then, the magnetic contribution can be inferred from the estimates of the first and second harmonic frequencies using Eq. (\ref{eq:varpiM}).

The associated uncertainty is determined by propagating the errors of all the other parameters whose estimates are inferred during the Bayesian analysis. As seen from Eqs. \eqref{eq:mu1mu2}, \eqref{eq:varpiM} and \eqref{eq:varpiGR}, the variance of the magnetic energy is related to the variance of $f_{(0)}$, $f_{(1)}$, and $\dot\varpi_{1\mathrm{PN}}$ such as $\sigma_{\dot{\varpi}_{\mathrm{M}}}^2 = \left(3 \pi \sigma_{f_{(0)}} / 2\right)^2 + \left(\pi \sigma_{f_{(1)}}\right)^2 + \left(3 \sigma_{\dot{\varpi}_{1\mathrm{PN}}}/2\right)^2$. The variance of the last term is much larger than the two others when eccentricity is present (i.e., $\sigma^2_{f_{(0)}} \ll \sigma_{\dot{\varpi}_{1\mathrm{PN}}}^2$ and $\sigma_{f_{(1)}}^2\ll \sigma_{\dot{\varpi}_{1\mathrm{PN}}}^2$) so that $\sigma^2_{\dot{\varpi}_{\mathrm{M}}}\simeq\sigma_{\dot{\varpi}_{1\mathrm{PN}}}^2$. From Eq.~(\ref{eq:mu1mu2}), we now deduce the uncertainty of the secular energy:
\begin{align}
  \left(\frac{\sigma_{\bar U_{\mathrm{M}}}}{\bar U_{\mathrm{M}}}\right)^2 &\simeq 3\left(\frac{\sigma_{\dot{\varpi}_{\mathrm{M}}}}{\dot{\varpi}_{\mathrm{M}}}\right)^2 + 3\left(\frac{\sigma_{\dot{f}_{(0)}}}{\dot{f}_{(0)}}\right)^2 + 12\left(\frac{\sigma_{f_{(0)}}}{f_{(0)}}\right)^2\nonumber \\
  &+\frac{169 \, e_0}{4}\left( \frac{\sigma_{e_0}}{e_0}\right)^2
\end{align}

The relative uncertainty $\sigma_{f_{(0)}}/f_0$ decreases from $10^{-10}$ after a 4-years LISA mission to $10^{-11}$ after a 8-years mission duration. The relative uncertainty  $\sigma_{\dot{f}_{(0)}}/\dot{f}_{(0)}$ decreases from $10^{-2}$ after a 4-years LISA mission to $10^{-3}$ after 8-years mission duration. However, the relative precision on the eccentricity is never better than $10^{-2}$ (the limit is the same after a 4-years or 8-years scenario; see Tab. \ref{tab:eccentricity}). Therefore, the uncertainty on the magnetic energy is essentially limited by $\sigma_{\dot{\varpi}_{\mathrm{M}}}$ which is itself limited by the uncertainty on the eccentricity and also on the uncertainty on the symmetric mass ratio of the binary system.  

\begin{figure}
    \centering
    \includegraphics[width=0.45\textwidth]{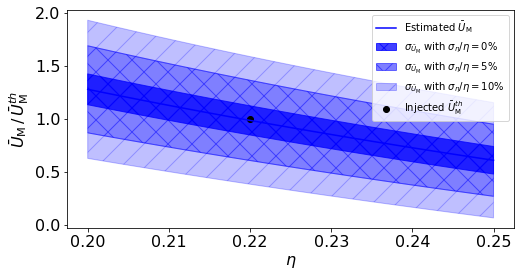}
    \caption{Estimate of the magnetic energy over its theoretical definition as a function of the symmetric mass ratio. The computation is realized for a 4-years LISA mission scenario. The \emph{black dot} represents the input value of the magnetic energy while the \emph{blue curve} is the estimate inferred from data analysis. The \emph{light blue area}, \emph{blue area} and \emph{dark blue area} represent respectively the magnetic energy uncertainty associated to a 10\%, 5\% and 0\% uncertainty levels on the knowledge of the symmetric mass ratio.} 
    \label{fig:MagneticExtraction}
\end{figure}

Because our transformation from the LDC parameters $(\mathcal{A}_{(0)},\mathcal{A}_{(1)},\iota,f_{(0)},f_{(1)},\dot{f}_{(0)},\dot{f}_{(1)})$ to the physical parameters $(\eta,e_0,n_0,\mathcal{M},D,\iota,\bar U_{\mathrm{M}})$ is under constrained, the magnetic secular energy is actually determined for different values of the symmetric mass ratio as discussed previously in Sect. \ref{sec:physfromGW}. Assuming that HM Cancri's symmetric mass ratio is exactly 0.22, namely $\sigma_{\eta}=0$, we see in Fig. \ref{fig:MagneticExtraction} (see the darkblue curve) that the magnetic energy is retrieved with a 20\% relative precision in the pessimistic case as summarized in Tab. \ref{tab:magnetic_energy}).

If we now assume that the symmetric mass ratio is not perfectly known, then it is likely that $\sigma_\eta$ will dominate in the computation of $\sigma_{{\bar{U}}_{\mathrm{M}}}$. As discussed previously in Sect.~\ref{sec:physfromGW}, we can guess \emph{an priori} value for the symmetric mass ratio, assuming that for typical GB, we expect $\eta\simeq 0.25$. However, in this case, the \emph{a priori} relative uncertainty on $\eta$ is expected to be relatively high, at least at the level of tenth of percent. The situation can be slightly improved relying on electromagnetic observations which could let to reach few percent level precision on the relative uncertainty of the symmetric mass ratio.

\begin{table}
	\centering
	\caption{HM Cancri magnetic energy (in $10^{36}$ J) estimates for different LISA mission duration scenarios for both optimistic and pessimistic cases assuming a known symmetric mass ratio (i.e., $\sigma_\eta=0$).}
	\begin{tabularx}{\linewidth}{c|>{\centering\arraybackslash}X >{\centering\arraybackslash}X}
	 \hline\hline
	 $\Big.$Case & Pessimistic & Optimistic \\
	 \hline
	4-years & $1.4 \pm 0.3$ & $1.4 \pm 0.2$ \\
	8-years & $1.4 \pm 0.2$ & $1.41 \pm 0.07$ \\
	\hline
	\end{tabularx}
	\label{tab:magnetic_energy}
\end{table}

Therefore, if we now consider that the symmetric mass ratio is known up to 5\% (resp. 10\%) precision, the estimate on the magnetic energy and associated uncertainty evolve as shown by the blue curve and the dark blue area (resp. light blue area) in Fig. \ref{fig:MagneticExtraction}.

All in all, these results suggest that the secular magnetic energy could indeed be determined with few percent relative precision ($5\%$ for the optimistic case) from the GW observations alone if the source is a strongly magnetic GB in quasi-circular orbit.

\section{Conclusion}
\label{sec:ccl}

In this work, we derived the waveform of the GW signal emitted by a strongly magnetic GB in quasi-circular orbit (up to second order in eccentricity). We considered the secular dynamics of the system assuming a dipole-dipole magnetic interaction and the GR point-mass interaction up to the 2.5PN approximation, according to results derived in \citet{PhysRevD.105.124042}. Using the LDC algorithms, we were able to simulate the GW signal of HM Cancri in the time domain and to generate the LISA response in the frequency domain. HM Cancri is a verification binary whose SNR is expected to be as high as 100 after 4-years of observation by LISA. In addition, it is a good eccentric candidate with an eccentricity about $0.1$, implying that the SNR of the second harmonic could be about 10 after a 4-years LISA mission. Moreover, it is made with two WDs whose magnetic fields could be as high as $10^9\ \mathrm{G}$. Then, by analyzing the signal in the frequency domain, we were able to show that the second harmonic is indeed visible after a 4-years LISA mission scenario, enabling for a determination of the eccentricity with a relative uncertainty at the level of 20\% and 5\% in the pessimistic and optimistic scenarios, respectively. These estimates, do not significantly change with the duration of the mission but with the \emph{a priori} knowledge of the inclination. Then, from the accurate determination of the frequencies of the first and second harmonics it is thus possible to estimate the secular magnetic energy of the binary system with a relative precision of 20\% percent after 4-years and 14\% after a 8-years LISA mission in the pessimistic case. The relative uncertainty on the magnetic energy can even be as low as $5\%$ after a 8-years LISA mission for the optimistic case.

If eccentricity and magnetism were not taken into account (as it is currently the case in the LDC), all catalogues of GBs would be biased for at least two reasons:
\begin{enumerate}
    \item The higher order harmonics would be considered as independent sources and would distort the count of the number of identified binaries. However, these falsely independent sources shall be identified with the same position in the sky so in theory it is possible to check if a signal could be a higher order harmonic of an existing source at the same sky localisation.
    \item The magnetic effect introduces a frequency bias that is indistinguishable from a binary with a slightly different frequency and then a different total mass. It would then be difficult for the community to draw conclusions about the process of formation or evolution of the GB as function of their masses.
\end{enumerate}

Until the launch of the mission, a new release of the Gaia catalogue will provide \emph{a priori} information for new verification binaries and thus simplify the search for these eccentric and magnetic binaries. In the mean time, the whole study is done assuming only patchy information about HM Cancri and should therefore be applicable to any GB with a sufficiently high SNR. It is expected that there will be binaries that are ``brighter'' than the verification binaries, for which the eccentricity and magnetic effect measurements will then be measurable with better accuracy.

The study carried out so far considers independently the main peak and the secondary peak of the GW and uses the LDC tools already developed to meet the specifications prior to the mission launch. This assumption, although simplistic, is sufficient to estimate, to within a few percent, the secular magnetic energy of the binary system. Although beyond the scope of this paper, it may be necessary to develop a tool to simultaneously search for the different harmonics in the GW signal in order to improve the accuracy on the eccentricity and magnetic effect retrieval.

\section*{Acknowledgments}
This work was supported by CNES, focused on LISA mission. C. Aykroyd acknowledges support from CNES and PSL/Observatoire de Paris for funding his PhD project. S. Mathis, C. Le Poncin-Lafitte and A. Bourgoin acknowledge support from PNPS of CNRS/INSU.

\section{Appendix : Amplitude and frequency uncertainties} \label{sec:appB}
In the main part of the paper, we assumed that the uncertainty in the frequency and amplitude of a GB depends mainly on the SNR ratio and the mission duration. To support this assertion analytically, we propose to demonstrate these dependencies using the same approach as the calculation done in the appendix of \citet{Savalle2022}. The Fisher matrix, $\Gamma^{\theta}_{ij}$ , provides an estimate of the precision to which the parameters $\theta$ of the source model can be determined from the data. Specifically, the uncertainty in parameter $\theta^i$, namely $\sigma_i$, can be estimated as $\sigma_i^2$ = $\left(\Gamma^{\theta}_{ii}\right)^{-1}$. It is defined as follows:  
\begin{equation}
    \Gamma^{\theta}_{ij} = 4 \Re{\left[\int_{0}^{\infty} \frac{(\frac{\partial  \tilde{s}(f)}{\partial \theta_i})^* \times (\frac{\partial  \tilde{s}(f)}{\partial \theta_i})}{S_n(f)}\mathrm{d}f\right]},
\end{equation}
where $\tilde{s}(f)$ is the Fourier transform of signal and $S_n(f)$ is the mission noise power spectral density.

For the sake of clarity, we will assume here that the GW signal can be modeled by a simple monochromatic oscillation at $f$ with amplitude $\mathcal{A}$ [cf. Eq. \eqref{eq:GB_WF}]. We assume that the signal is observed for a time T corresponding to the duration of the mission. The response of the LISA interferometer is neglected here since it only brings corrections of higher orders to the computation detailed hereafter.

The time domain form of the signal is therefore :
\begin{equation}
\begin{aligned}
    s(t) &\equiv \mathcal{A} \cos(2 \pi f t + \phi_0) \times \Pi_T(t-T/2) \\
    &= h_0(t) \times \Pi_T(t-T/2),
\end{aligned}
\end{equation}
where $\phi_0$ is an arbitrary phase and $\Pi_T$ is the rectangular function such that :
\begin{equation}
    \Pi_T(t) = \left\{ \begin{array}{c c c} 0 &\text{if} & t > T/2 \\ 1/2 &\text{if} & t = T/2 \\ 1 &\text{if} & t < T/2 \end{array}\right.\,.
\end{equation}
The frequency domain form of the signal is :
\begin{equation}
\begin{aligned}
    \tilde{s}(f) &= \tilde{h}(f) \ast \tilde{\Pi_T}(f)
    = \mathcal{A} \pi T \text{sinc}(\pi f T).    
\end{aligned}
\end{equation}

In order to estimate the uncertainty on the frequency and amplitude, we need to calculate the derivatives of the Fourier transform of the signal with respect to these two parameters :
\begin{equation}
\begin{aligned}
\frac{1}{\tilde{s}} \frac{\partial \tilde{s}}{\partial f} &= \frac{1}{f}\left(\pi f T \cot{(\pi f T)} -1\right) \\
\frac{1}{\tilde{s}} \frac{\partial \tilde{s}}{\partial {\mathcal A}} &= \frac{1}{\mathcal A}.
\end{aligned}
\end{equation}

The diagonal element of the Fisher matrix are :
\begin{equation}
\begin{aligned}
    \Gamma_{f f} & \simeq \left(\frac{1}{f}\left(\pi f T \cot{(\pi f T)} -1\right)\right)^2 \times \textsc{SNR}^2
    \\
    \Gamma_{\mathcal{A} \mathcal{A}} &\simeq \frac{1}{\mathcal{A}^2} \times \textsc{SNR}^2,
\end{aligned}
\end{equation}
where the SNR is defined as :
\begin{equation}
    \textsc{SNR}^2 = 4 \Re{\left[\int_{0}^{\infty} \frac{\tilde{s}(f)^* \times \tilde{s}(f)}{S_n(f)}\mathrm{d}f\right]}.
\end{equation}

Noticing that the cotangent function is bounded by 2, that the product $f \times T$ is large in front of 1, and recalling that the uncertainty of a parameter is the square root of the inverse of its element in the Fisher matrix, we obtain: 
\begin{equation}
        \sigma_{f} \propto \frac{1}{\textsc{SNR}} \frac{1}{T} \, , \qquad
        \frac{\sigma_{\mathcal{A}}}{\mathcal{A}} \propto \frac{1}{\textsc{SNR}}.
\end{equation}

\bibliographystyle{apsrev4-1}
\bibliography{GW_eccentric}

\end{document}